# $Mg_{3+\delta}Sb_xBi_{2-x}$ family: A promising substitute for the start-of-art n-type thermoelectric materials near room temperature

Rui Shu[1], Yecheng Zhou[2], Qi Wang[1], Zhijia Han[1], Yongbin Zhu[1], Yong Liu[1,3], Yuexing Chen[2,6], Meng Gu[1], Wei Xu[4], Yu Wang[5], Wenqing Zhang[2], Li Huang[2*] and Weishu Liu[1*]

[1]Department of Materials Science and Engineering, Southern University of Science and Technology, Shenzhen 518055, China

[2]Department of Physics, Southern University of Science and Technology, Shenzhen 518055, China

[3]Beijing Institute of Aeronautical Materials, AECC, Beijing 100095, China

[4]Beijing Synchrotron Radiation Facility, Institute of High Energy Physics, Chinese Academy of Sciences, Beijing 100049, China

[5]Shanghai Synchrotron Radiation Facility, Shanghai Institute of Applied Physics, Chinese Academy of Sciences, Shanghai 201204, China

[6]Shenzhen Key Laboratory of Advanced Thin Films and Applications, College of Physics and Energy, Shenzhen University, Shenzhen 518060, China

*E-mail: huangl@sustc.edu.cn, liuws@sustc.edu.cn

## Abstract

$Bi_2Te_{3-x}Se_x$ family has been the n-type start-of-the-art thermoelectric materials near room temperatures (RT) for more than half-century, which dominates the active cooling and novel waves harvesting application near RT. However, the drawbacks of brittle nature and Te-containing restrict the further applications exploring. Here, we show that a $Mg_{3+\delta}Sb_xBi_{2-x}$ family ($(ZT)_{avg}$ =1.05) could be a promising substitute for the $Bi_2Te_{3-x}Se_x$ family ($(ZT)_{avg}$ =0.9-1.0) in the temperature range of 50-250 °C based on the comparable thermoelectric performance through a synergistic effect from the tunable band gap using the alloy effect and the suppressible Mg-vacancy formation using interstitial Mn dopant. The former is to shift the optimal thermoelectric performance to near RT, and latter is helpful to partially decouple the electrical transport and thermal transport in order to get an optimal RT power factor. A positive temperature-dependence of band gap suggested this family is also a superior medium-temperature thermoelectric material for the significantly suppressed bipolar effect. Furthermore, a two times higher mechanical toughness, compared with $Bi_2Te_{3-x}Se_x$ family, consolidates the promising substitute for the start-of-art n-type thermoelectric materials near RT.

## 1. Introduction

Thermoelectric devices attracted increasing attention from both the academic and industrial communities because they offer a promising solution for convenient conversion between heat and electricity.[1] The power generation applications, such as solid state solar thermoelectric generator and waste heat harvester from heat sources of automotive exhaust gas, fireplace and steel industrial, gravitated intensively academic efforts on the medium temperature thermoelectric materials.[2] On the other hand, the industrial demands in the thermal management of microelectronic integrated circuits, photonic LED packaging, and the self-powered system for the IoT sensors, also continuously encouraged the works on the room-temperature thermoelectric materials.[3] With continuously efforts since 1950s, $Bi_2Te_3$ and its alloys have been the state-of-the-art thermoelectric materials near RT and hence a long history as commercially available solutions for silent cooling. The *ZT* values of 1.1 - 1.2 at 357 K for n-type[4, 5] and ~1.4 at 373 K for p-type[6, 7] were widely reported in the $Bi_2Te_3$-based alloys. Here, the symbol *ZT* is the thermoelectric performance scale, defined as $ZT = TS^2\sigma/\kappa$, where $T$, $S$, $\sigma$ and $\kappa$ are the average temperature, Seebeck coefficient, electrical conductivity and thermal conductivity, respectively. However, there are well-known disadvantages in the $Bi_2Te_3$-based materials, i.e. brittle and Te-contained. Tellurium is scare in earth's crust with a mass abundance of $1\times10^{-3}$ ppm, less than Gold ($4 \times 10^{-3}$ ppm)[8], which would limit application of the $Bi_2Te_3$-based devices. There was a long list on the novel thermoelectric materials since 1990s, such as Skutterudite[9], $Zn_4Sb_3$[10], $In_4Se_{3-\delta}$[11], $Cu_2CdSnSe_4$[12], $CuSe_2$[13], MgAgSb[14], SnTe[15], BiCuSeO[16], SnSe[17], etc. Among them, only p-type MgAgSb showed a promising as new candidate for the thermoelectric application near RT. Unfortunately, it still contained 42 wt.% noble metal silver, which was definitely not beneficial to the large scale application. One of the authors' early efforts was the compound of $Mg_2Sn_{0.75}Ge_{0.25}$, with *ZT* values of 0.4 at RT and an increased value up to 1.0 at 250 °C[18], which was still not good enough to compete the n-type $Bi_2Te_{2.7}Se_{0.3}$ within the considered temperature of RT-250 °C.

Recently, the Zintl compounds got widely attentions due to its big family members and also high elemental abundancy. In 2016, Tamaki *et al.*[19] reported that an $Mg_{3+\delta}Sb_{1.5}Bi_{0.5}$ had a *ZT* values of 1.5 near 450 °C, which immediately gravitated many efforts to push up the peak *ZT* in medium temperature rang. Some of strategies including defect chemistry[19, 20], the carrier scattering mechanism manipulation[21, 22], ground boundary engineers[23], have successfully performed to increase the peak *ZT* from 1.5 up to 1.7 near 500 °C. Here, in contract, the boost of thermoelectric performance near RT attracted us more. Promisingly, Nb-doped $Mg_{3+\delta}Sb_{1.5}Bi_{0.5}$ compound[21] has achieved a double *ZT* value (0.4) at RT compared with that (*ZT* = 0.2) of Tamaki's work[19].

In this study, we will show that n-type $Mg_{3+\delta}Sb_xBi_{2-x}$ Zintl family has promising to substitute for the n-type start-of-the-art thermoelectric materials $Bi_2Te_{2.7}Se_{0.3}$ (ingot: $(ZT)_{avg}$ = 0.89; Nano: $(ZT)_{avg}$ = 1.01) due to the comparable average *ZT* ($(ZT)_{avg}$ = 1.05) in the temperature range of 50-250 °C and also a significantly superior mechanical toughness ($K_{IC}$=3.0 MPa·m$^{1/2}$) and low cost. The enhanced thermoelectric and mechanical performance were achieved via carefully tuning band gap by the alloy effect, successfully suppressing the formation of Mg vacancies through Mn dopant, and nanostructured grain boundary.

## 2. Results and discussions

We will first uncover the process how we got the breakthrough to identify this new RT thermoelectric material. The theoretical predication of band gap for $Mg_3Sb_xBi_{2-x}$ family linear increased from -0.02 eV to 0.60 eV as the Sb content changes from x = 0.5 to x = 2, indicating the transition from metal to semiconductor.[24, 25] The conduction band bottom was formed by the $CB_1$-band between the M* (0, 0.417, 0) and L* (0, 0.417, 0.5) coordinate, while the valence top was at the Γ point. The compositional resolution of band-gap was mainly ascribed to the shift of $CB_1$ band with the varying ratio of Sb/Bi, as schematically shown in **Figure 1**a-b. According to the empirical trend of band gap ($E_g$) dependent application temperature range, the room-temperature TE materials has smaller band gap than that of medium temperature and high-temperature TE materials. Pure $Bi_2Te_3$ usually has a band gap of $E_g$ = 0.145 eV and a slightly increased band gap due to alloy effect of Se, i.e. $E_g \sim (0.145+0.049\cdot x)$ eV for the general formula of $Bi_2Te_{3-x}Se_x$ when $x < 0.9$.[26] Therefore, the band gap of classic $Bi_2Te_{2.7}Se_{0.3}$ gave us the hint to choose the composition range from x = 1.1 to x = 1.5 in the $Mg_3Sb_xBi_{2-x}$ family, as shown in **Figure 1**c-d. Fourier transform infrared (FTIR) spectroscopy measurements with Kramers–Kronig analysis were conducted for the $Mg_3Sb_xBi_{2-x}$ family (x = 1.1, 1.2, 1.3, 1.4 and 1.5). It was found the RT-band gap of $Mg_3Sb_xBi_{2-x}$ family is 0.242 eV, 0.245 eV, 0.249 eV, 0.268 eV and 0.279 eV for the x = 1.1, 1.2, 1.3, 1.4 and 1.5, respectively. The measured optical band gaps were slightly smaller than theoretical calculation values, and the Sb/Bi ratio dependent band gap change was confirmed.

Furthermore, Mg vacancies in $Mg_3Sb_xBi_{2-x}$ family were considered as one of major reasons for the abnormal thermoelectric transport behavior near RT in recent reports [19-22]. It is reported that Cu atom went into the interstitial site of Van de Waals layer in $Bi_2Te_3$ and significantly suppressed the formation of Te vacancy[4]. Here, a scenario for the suppressible Mg vacancies in Mn doped $Mg_3Sb_xBi_{2-x}$ were proposed from first-principle calculations and confirmed by experimental techniques. The crystal structure of $Mg_3Sb_2$ is a laminar structure with alternately $[Mg_2Sb_2]^{2-}$ layers and $[Mg]^{2+}$ layers. The Mg atom in the $[Mg]^{2+}$ layers is also labeled as Mg1 while the Mg atom in $[Mg_2Sb_2]^{2-}$ layers is denoted as Mg2. The formation energies of Mg vacancies in $Mg_3Sb_2$ structure were investigated with first-principles calculations. It was shown than the Mn dopant would reduce Mg1 and Mg2 vacancies in n-type $Mg_3Sb_2$ and even in weak p-type conditions. **Figure 2**a gives a schematic figure of the mechanism of Mn dopant reducing the Mg vacancy. In **Figure 2**b, considered Fermi energy ($E_F$) was located at the conduction band minimum (CBM), the formation energies of vacancy at Mg1-site increased from 0.61 eV to 0.92 eV with a Mn atom taking the interstitial site, while the value of the Mg2-site vacancy raised from 0.55 eV to 0.82 eV, which indicated that the Mn doping could help stabilize Mg atom via suppressing its evaporation during sintering process, and finally reduce the concentration of Mg vacancies in final obtained bulk materials. Experimentally, the Mn *K*-edge XAFS spectra was measured for as-fabricated $Mg_{3+\delta}Sb_{1.5}Bi_{1.49}Te_{0.01}$: $Mn_{0.01}$ sample to study the most probable Mn defects (XAFS: X-ray Absorption Fine Structure Spectroscopy). The simulation of Mn *K*-edge XAFS spectra was conducted on the basis of full multiple scattering (FMS) theory within muffin-tin approximation by using a theoretically optimized crystal structure. According to the main feature of the measured XAFS spectra in **Figure 2**c, the as-fabricated Mn *K*-edge displayed a character peak at 6450 eV and 6555 eV. Among the five most likely Mn positions, *i.e.* interstitial sites: IA (2/3, 1/3, 0.182; or 1/3, 2/3, 0.581) and IC (0, 1/2, 1/2); substitution-sites: Mg1(0, 0, 0), Mg2(1/3, 2/3, 0.631; or 2/3, 1/3, 0.368) and Sb (1/3, 2/3, 0.225; or 2/3 1/3 0.775), only

the simulated spectra of IA showed the consistent features with experimental measurement, which further confirmed that IA site, located at the interstitial layer between the $[Mg_2Sb_2]^{2-}$ layer and $[Mg]^{2+}$ layer, was the most probable site for the Mn dopant.

To investigate the thermoelectric transport properties of the $Mg_3Sb_xBi_{2-x}$ family, the nominal chemical composition we used was $Mg_{3+\delta}Sb_xBi_{2-x-y}Te_y$: $Mn_z$. Here, extra Mg ($\delta$ = 0.2) were used to compensate the loss according to the references[21, 22], and extra Mn (z=0.01) was used to reduce Mg vacancy in this study. Noticed that the recent study on anisotropy of $Mg_3Sb_2$ compounds[27], electrical and thermal transport were both measured in directions perpendicular to press direction. No obvious impurity phase was observed from the XRD measurement, which was given in **Figure S1**. Tellurium substitution (y=0.01) at the anion-site was chosen as the carrier donor. The thermoelectric properties of the classical composition $Mg_{3.2}Sb_{1.5}Bi_{0.49}Te_{0.01}$ and $Mg_{3.2}Sb_{1.99}Te_{0.01}$ were firstly double checked in this work and compared with the previously reported value[21, 28, 29], as illustrated in **Figure 3**a-c. The as-fabricated $Mg_{3.2}Sb_{1.5}Bi_{0.49}Te_{0.01}$ showed a good consistence with previous reports[21, 28], while the as-fabricated $Mg_{3.2}Sb_{1.99}Te_{0.01}$ presented slightly higher power factor than that of the Ohno's work[29] which might be resulted from the difference in fabrication processes. Furthermore, a negative temperature dependence of electrical resistivity was confirmed in these two Mn-free samples, which is well consistent with the reported results [21, 28, 29]. **Figure 3**d-f shows the thermoelectric transport properties of $Mg_{3+\delta}Sb_xBi_{1.99-x}Te_{0.01}$: $Mn_{0.01}$ family (x = 1.1, 1.2, 1.3, 1.4 and 1.5). Obviously, the temperature dependent electrical resistivity of all the Mn-contained samples displayed a metallic behavior. No abnormal ionic impurity scattering was observed in all the Mn-contained samples. Furthermore, the electrical resistivity and Seebeck coefficient of considered $Mg_{3+\delta}Sb_xBi_{1.99-x}Te_{0.01}$: $Mn_{0.01}$ family presented a weak relation with the ratio of Bi/Sb, with average values of 15.0 μΩ m and 175.0 μV $K^{-1}$ and hence a high power factor of over 2000 μW $m^{-1}$ $K^{-2}$ near RT. Among the samples, the sample with Sb content of x=1.5 showed the highest power factor of 2684 μW $m^{-1}$ $K^{-2}$ at 150 °C, which is higher than the previously reported PFs, such as, Nd-doped $Mg_{3.2}Sb_{1.5}Bi_{0.5}$ (~2059 μW $m^{-1}$ $K^{-2}$ at 150 °C[21]) and Co-doped $Mg_{3.2}Sb_{1.5}Bi_{0.5}$ (~2200 μW $m^{-1}$ $K^{-2}$ at ~230 °C[22]).

**Figure 4**a shows the thermal conductivity of $Mg_{3+\delta}Sb_xBi_{1.99-x}Te_{0.01}$: $Mn_{0.01}$ family, which shows a weak dependence on the Sb content within the considered temperature range. At RT, the average thermal conductivity is 1.12 W $m^{-1}$ $K^{-1}$ which is slightly higher than that of Nd-doped $Mg_{3.2}Sb_{1.5}Bi_{0.5}$[21] and Co-doped $Mg_{3.2}Sb_{1.5}Bi_{0.5}$[22]. Temperature dependence of thermal diffusivity and specific heat of as-fabricated $Mg_{3+\delta}Sb_xBi_{1.99-x}Te_{0.01}$: $Mn_{0.01}$ were provided in **Figure S2**. The *ZT* value of the $Mg_{3+\delta}Sb_xBi_{1.99-x}Te_{0.01}$: $Mn_{0.01}$ family was calculated from the measured electrical resistivity, thermal conductivity, as well as the Seebeck coefficient, and plotted in **Figure 4**b. *ZT* values of 0.50-0.63 at RT were observed in as-fabricated $Mg_{3+\delta}Sb_xBi_{1.99-}$

$_xTe_{0.01}$: $Mn_{0.01}$ family, which showed weak dependence on the Sb/Bi ratio. Furthermore, the $Mg_{3+\delta}Sb_xBi_{1.99-x}Te_{0.01}$: $Mn_{0.01}$ family possessed a negligible bipolar effect according to the temperature dependent thermal conductivity and Seebeck coefficient, even the one (x=1.1) having the smallest band gap of 0.24 eV at RT. As a result, high *ZT* values of 1.5-1.7 were reached for all the family, which was consistent with the recent reports $Mg_{3+\delta}Sb_{1.5}Bi_{0.5}$ [21, 22, 24]. However, the suppressed bipolar effect initiated our interesting to measure the temperature dependence of optical band gap, i.e. $dE_g/dT$. **Figure 4**c depicts the temperature dependent band gap of as-fabricated $Mg_{3+\delta}Sb_xBi_{1.99-x}Te_{0.01}$: $Mn_{0.01}$ family, details optical spectra information and gap values are presented in **Figure S3** and **Table S1**. Surprising that as-fabricated $Mg_{3+\delta}Sb_xBi_{1.99-x}Te_{0.01}$: $Mn_{0.01}$ family showed a positive $dE_g/dT$ value in **Figure 4**d, similar to the PbX(Te, Se,X)[30], $(Hg_{1-x}Cd_x)Te$[31] and $Cd_3As_2$[32]. In contrast, most of the known thermoelectric materials, including $Bi_2Te_{3-x}Se_x$, $Mg_2X$ (X=Sn, Ge, Si) and $CoSb_3$, possesses a negative $dE_g/dT$ value. The normally negative $dE_g/dT$ of band gap is understood from the thermal expansion effect, *i.e.* the term of $(dE_g/dT)_{epn}$ in the Eq. (1). Additionally, the electron-phonon interaction also could contribute the temperature dependence of band gap, i.e. $(dE_g/dT)_{el\text{-}ph}$ which could be a negative term[33]. As a result, an abnormally positive $dE_g/dT$ value could be presented. Although the clear physical picture for the relation between the negative $dE_g/dT$ and the electron-phonon interaction has still remained unknown, it did not deflate enthusiasm to explore this material family for thermoelectric application.

$$\frac{dE_g}{dT} = \left(\frac{dE_g}{dT}\right)_{epn} + \left(\frac{dE_g}{dT}\right)_{el-ph} \quad (1)$$

Among the family, the excellent reproducibility in thermoelectric properties of $Mg_{3+\delta}Sb_{1.5}Bi_{0.5}$ and $Mg_{3+\delta}Sb_{1.2}Bi_{0.8}$ were given in **Figure S4-5**. Moreover, the highest performance also showed a good thermal stability upon temperature changes (**Figure S6**).

**Figure 5**a compares the thermoelectric *ZT* values of as-fabricated n-type $Mg_{3+\delta}Sb_xBi_{2-x}$ (x=1.2 and 1.5) with the state-of-the-art n-type $Bi_2Te_{3-x}Se_x$ family thermoelectric materials and other early efforts for the room temperature application, meanwhile other thermoelectric transport properties can be seen in **Figure S7**. $Bi_2Te_{3-x}Se_x$ family overwhelmed the known n-type materials a long time in the considered temperature range of RT-250 °C. Particularly, the temperature dependent *ZT* values of a commercial ingot and a nanostructured $Bi_2Te_{3-x}Se_x$ [4] were presented for the benchmark lines, which show typical *ZT* values of 0.7-0.9 at RT, respectively. Near RT, we show that a *ZT* value of 0.65 was achieved by carefully tuning band gap, the point defects and Mn dopant. For more accurate comparison of thermoelectric performance in a temperature range, the average $(ZT)_{avg}$ and engineering figure of merit $(ZT)_{eng}$, as defined in Eq.(2-3)[34], were calculated. The detailed relationships between $(ZT)_{avg}$, $(ZT)_{eng}$ and leg efficiency were given in Supporting Information.

$$(ZT)_{avg} = \frac{1}{T_h - T_c} \int_{Tc}^{Th} ZT(T)\, dT \qquad (2)$$

$$(ZT)_{eng} = \frac{\left(\int_{Tc}^{Th} S(T)\, dT\right)^2}{\left(\int_{Tc}^{Th} \rho(T)\, dT\right)\left(\int_{Tc}^{Th} \kappa(T)\, dT\right)} (T_h - T_c) \qquad (3)$$

**Figure 5**b-c shows the calculated $(ZT)_{avg}$ and $(ZT)_{eng}$ with in the temperature range of 50-250 °C. The $Mg_{3+\delta}Sb_xBi_{2-x}$ family in this work showed $(ZT)_{avg}$ values of 1.03 (x=1.3) and 1.05 (x=1.5), $(ZT)_{eng}$ values of 0.48 (x=1.3) and 0.49 (x=1.5) which were higher than that of both the commercial $Bi_2Te_{3-x}Se_x$ ingot ($(ZT)_{avg}$ = 0.89, $(ZT)_{eng}$=0.42) and a nanostructured $Bi_2Te_{3-x}Se_x$ ($(ZT)_{avg}$ =1.01, $(ZT)_{eng}$=0.47)[5]. It was also superior as compared with both $Mg_2Sn_{0.75}Ge_{0.25}$ ($(ZT)_{avg}$ = 0.32, $(ZT)_{eng}$=0.69)[18] and Nb doped $Mg_{3+\delta}Sb_{1.5}Bi_{0.5}$ ($(ZT)_{avg}$ = 0.42, $(ZT)_{eng}$ = 0.90)[21]. This result suggested that the $Mg_{2+\delta}Sb_xBi_{2-x}$ family could be a superior candidate for the application in the power generation application during the temperature range of RT-250 °C, such the solar thermoelectric power generation [35].

**Figure 6**a depicts the fracture toughness $K_{IC}$ and Young's modulus $E$ of as fabricated n-type $Mg_{3+\delta}Sb_xBi_{2-x}$ family, together with other thermoelectric materials: $Bi_2Te_3$[36, 37], PbTe[38], MgAgSb[39], $CoSb_3$[40], SiGe[41] and $CaMnO_3$[42]. Firstly, the Young's modulus of $Mg_{3+\delta}Sb_xBi_{2-x}$ family (42-45 GPa) was comparable with that of $Bi_2Te_3$ family (40-45 GPa) and PbTe (51-77 GPa), MgAgSb (55 GPa), while smaller than that of $CoSb_3$(118-143 GPa), SiGe (119-135 GPa) and $CaMnO_3$ (142-160 GPa). $Bi_2Te_3$-based thermoelectric material is notorious for its brittle nature in industry. The typical $K_{IC}$ value of commercial $Bi_2Te_3$-ingot is around 0.6-0.7 MPa·m$^{1/2}$. Our measurements showed that a p-type $Bi_{2-x}Sb_xTe_3$ presented a $K_{IC}$ value of 0.66 MPa m$^{1/2}$ and $E$ value of 51 GPa. Zhao *et al.* [36] reported the $K_{IC}$ value was in the range of 1.15-1.35 MPa·m$^{1/2}$ for the n-type nanostructured $Bi_2Te_3$ and a weak dependence on the addition of SiC nano particles. Later, Liu *et al.* [43] confirmed the addition of SiC nano particles improved the mechanical properties of p-type $Bi_{0.5}Sb_{1.5}Te_3$, and showed an increased modulus value from 40.1 to 44.9 GPa and $K_{IC}$ value from 0.82 to 0.91 MPa·m$^{1/2}$ with the addition of 1% SiC nano particles. The comparison tests of the fracture toughness (0.89 MPa·m$^{1/2}$) and modulus (49.0 GPa) for n-type nanostructured $Bi_2Te_{2.7}Se_{0.3}$ in this work showed a good agreement. Although the nanostructure engineering enhanced the mechanical properties of $Bi_2Te_3$ family to some extent, it was still far from changing the brittle nature. In this work, the $Mg_{3+\delta}Sb_xBi_{2-x}$ family gives chances to improve the mechanical performance for RT thermoelectric application. It showed a much higher $K_{IC}$ value in the range of 2.2-3.0 MPa·m$^{1/2}$, overwhelmed most of the known thermoelectric materials, such as PbTe (0.34 MPa·m$^{1/2}$)[38], Skutterudites (1.5 MPa·m$^{1/2}$)[44] and SiGe (1.0 MPa·m$^{1/2}$)[41]. Furthermore, in **Figure 6**b, the $K_{IC}$ value of $Mg_{3+\delta}Sb_xBi_{2-x}$ family showed a notable dependence on the Sb/Bi ratio. The $K_{IC}$ value increased from 2.20 to 3.03 MPa·m$^{1/2}$ as the Sb content increased form x=1.1 to 1.3. While for the one with Sb content of x>1.3, the $K_{IC}$ value of $Mg_{3+\delta}Sb_xBi_{2-x}$ family was saturated, with a value of 3.0 MPa·m$^{1/2}$. For further verification of the good toughness

of $Mg_{3+\delta}Sb_xBi_{2-x}$ family, the inset in **Figure 6**b, a Vickers indentation formed under the load of 20 N, presented the intriguing crack patterns on the bulk surfaces of $Mg_{3+\delta}Sb_{1.2}Bi_{0.8}$ sample. It was characterized by the cracks flaking off around the indentation partly, instead of converging together and emanating outside. The good mechanical toughness further suggests adopting the $Mg_{3.2}Sb_xBi_{2-x}$ family to be a new candidate for the thermoelectric application near RT.

**Figure 7**a shows a typical morphology of the fracture surface of as-fabricated $Mg_{3+\delta}Sb_xBi_{1.99-x}Te_{0.01}$: $Mn_{0.01}$(x=1.2). It was found that most of grains exhibited the transgranular fracture mode, with an average grain size larger than 5 μm. The coarse grain was also confirmed by the low-magnification TEM bright-field image in **Figure 7**b-c. Note that, the grain size of as-fabricated $Mg_{3+\delta}Sb_xBi_{1.99-x}Te_{0.01}$: $Mn_{0.01}$ family was bigger than the recent reports about n-type $Mg_3Sb_{1.5}Bi_{0.5}$ prepared by 2-minute hot pressing at 650 °C[21]. The large grain sizes are favorable for reducing the electrical resistivity due to less grain boundary scattering[23, 28] and could be an important factor to shape the high power factor of as-fabricated $Mg_{3+\delta}Sb_xBi_{1.99-x}Te_{0.01}$: $Mn_{0.01}$ family, which was further verified by an another recent report about Mn-doped $Mg_{3.2}Sb_{1.5}Bi_{0.5}$[45] where even though manganese dopant used, no extraordinary thermoelectric performance was observed near RT due to the smaller grain size (~200 nm). The selected area electron diffraction (SAED) pattern (inset in **Figure 7**c) of the $Mg_{3+\delta}Sb_xBi_{1.99-x}Te_{0.01}$: $Mn_{0.01}$ (x=1.2), which confirmed the single-phase of $La_2O_3$-type

trigonal structure (Space group: P-3m1, 164). The high resolution transmission electron microscope image displayed some interesting nano features, including nano precipitations and nanostructured grain boundaries (GB), which might be in favor of the low thermal conductivity. The nano precipitations were widely observed in thermoelectric materials synthesized by metallurgy route, such as (Bi, Sb)$_2$Te$_3$[7], PbTe[46], half-Heusler compounds[47]. Shuai *et al.*[48] found the trace of Mg precipitation formed at the grain boundary of $Mg_{3+\delta}Sb_{1.5}Bi_{0.49}Te_{0.01}$. The similar microstructure scenario was not found in this work. In our case, the nano inclusion should be due to the local composition fluctuation between Sb and Bi. **Figure 7**d shows an interesting nanostructured grain boundary (GB), which a common feature for this material system. The inset of **Figure 7d** shows that the observed grain boundary was gathering with (002)-plane and (112)-plane $Mg_3Sb_2$ crystals, suggesting a stress related contrast rather than precipitations. Recently, we noted that dense dislocations could be formed at the grain boundary in both the $Bi_{0.5}Sb_{1.5}Te_3$[49] and $Yb_xCo_4Sb_{12}$[50] by adding extra low melting metal Te and Sb during the sintering process, respectively. We speculated that the observed nanostructured GB would be related to the extra Mg. However, the morphology of the grain boundary could significantly change with content of extra Mg and also the sintering process, which was worthy of a systematic investigation in future.

For the potential room temperature application, the estimated cost of as-fabricated $Mg_{3+\delta}Sb_xBi_{2-x}$ is between 9.5 to 11.3 $/kg according to the US Geological Survey Data

Series (2015) [51], shown in **Table S2**. It is only about quarter price of the classic n-type $Bi_2Te_{2.7}Se_{0.3}$ (44.1 $/kg), and comparable to the conventional n-type $CoSb_3$ and $Mg_2Si$, indicative of a promising large industrial use as thermoelectric devices for low-cost requirement.

## 3. Conclusion

In this work, we have experimentally showed that the $Mg_{3+\delta}Sb_{2-x}Bi_x$ family had a high comparable thermoelectric performance in terms of the average *ZT* or engineering *ZT* in the temperature of 50-250 °C, but two times higher toughness and only half cost as compared with the n-type state-of-the-art room temperature thermoelectric materials $Bi_2Te_{2.7}Se_{0.3}$. The significantly boost of room-temperature thermoelectric performance was synergistically effect of carefully tuning band gap through the alloy effect and successfully suppressing the formation of Mg vacancy through a Mn dopant. We also for the first time observed a positive temperature-dependent band gap of $Mg_{3+\delta}Sb_{2-x}Bi_x$ Zintl family, which directly supported for its suppressed bipolar effect in the medium temperature range and hence a high peak ZT of 1.5-1.6 at 500 °C. The coarse-grained boundaries were observed which could positive contribute the high fracture toughness about 3.0 $MPa\cdot m^{1/2}$. Additionally, an unprecedented mechanical toughness, in terms of the industrial application, gives a more reliable alternative to commercial n-type thermoelectric materials in the temperature of RT-250 °C.

## 4. Experimental section

### Synthesis

High purity magnesium turnings (Mg, 99.8%; Alfa Aesar), antimony shots (Sb, 99.999%; 5N Plus), bismuth shots (Bi, 99.999%; 5N Plus), tellurium shots (Te, 99.999%; 5N Plus), and manganese powers (Mn, 99.95%; Alfa Aesar) were weighed according to the composition of $Mg_{3+\delta}Sb_{2-x}Bi_x$ ($\delta$=0.02, x=1.1, 1.2, 1.3, 1.4, and 1.5). The elements were loaded into a stainless-steel ball milling jar in a glove box under an argon atmosphere with the oxygen level (<0.1 ppm). The materials were ball-milled for 10 hours. The ball-milled powders were then loaded into a graphite die of the inner diameter of 15 mm in a glove box. The graphite die with loaded powder was immediately hot pressed at 675 °C under a pressure of 50 MPa for 5 min. The thickness of hot-pressed disks was about 10 mm, which allow to measure Seebeck coefficient and electrical conductivity and thermal conductivity in the same direction, i.e. the directions perpendicular to pressure.

### Structural characterization

*X-Ray Diffraction (XRD)*

The powder diffraction patterns of the various samples were obtained using a Rigaku Smartlab diffractometer with Cu K$\alpha$ radiation ($\lambda$ = 1.5406 Å operating at 40 kV/15 mA with a $K_\beta$ foil filter.

*X-ray Absorption Fine Structure spectroscopy (XAFS)*

The XAFS spectra at Mn K-edge was measured at the BL14W beamline of Shanghai Synchrotron Radiation Facility. The spectra of all samples are collected in Fluorescence mode using the Si (111) double crystal monochromator, gas filled ion chamber and Lytle-type detector. All raw data were preprocessed using the IFEFFIT package[52] following the conventional procedure: normalization to unit edge jump after removal of atomic absorption background mimicked using the Autobk algorithm implemented in the Athena code. The simulation was implemented in FEFF9 code[53], on the basis of the Full Multiple Scattering (FMS) theory within muffin-tin approximation. The Hedin-Lundqvist correlation potential were adopted and the atomic scattering potential were calculated with the self-consistent field (SCF) method. The radius of the atomic clusters for FMS and SCF calculations are selected as 10 and 6 Å, respectively.

*Transmission Electron Microscopy (TEM)*

TEM was conducted with FEI TECNAI F30 with an operating voltage of 300kV. The specimen for HRTEM examination was made by mechanical grinding, followed by $Ar^+$-ion milling, which was performed on a Gatan 691 Precision Ion Polishing Systems with liquid nitrogen cooling to -173 °C. Initially, the incident angle and energy of $Ar^+$ ions were set at 6° and 3.2 KeV, respectively. During the final step, they were reduced down to 2° and 0.5 KeV, respectively.

**Thermoelectric characterization**

All the samples were cut into about 2.5 mm × 3 mm × 14 mm pieces, of which the non-contact three faces were coated with a thin-layer BN to protect instruments, to simultaneously measure the electrical resistivity and Seebeck coefficient under a low-pressure helium atmosphere from RT to 500 °C (ZEM-3; ULVAC Riko). Thermal conductivity, $\kappa = dDC_p$ was calculated using the measured density (d) by the Archimedean method, thermal diffusivity (D, square simples in 6×6×1 $mm^3$) by the laser flash method (LFA 467; Netzsch), and specific heat (Cp) by differential scanning calorimetry (DSC 404C; Netzsch, details are explained in the Supporting Information).

**Band gap measurement**

Fourier transform infrared (FTIR) spectroscopy was conducted in the mid-IR range (4000 ~ 400 $cm^{-1}$) on an infrared spectrometry (IS50, Thermo Nicolet) between RT to 300 °C. The temperature-dependence band gaps were derived based on the Kramers-Kronig equation: $\alpha/S = (1-R)^2/(2R)$, where $R$ is the reflectance, α and $S$ are the absorption and scattering coefficients, respectively.

**Mechanical properties measurement**

The density of the samples was measured by the Archimedes method. Fracture toughness ($K_{Ic}$) was measured using the three-point flexural specimens by the single-edge notch beam test (ElectroForce 3200, a computer-controlled testing machine), testing machine. The tested specimens were cut to the same nominal dimensions 14×3

×1.5 mm by wire electrical discharge machining. The starting notch with a length of half the width was created at the midspan of the sample, and the length of pre-crack was less than 0.1 mm. the span was set at 8 mm, and the static load was conducted at a cross-head speed of 0.01 mm/min until complete failure. The $K_{IC}$ can be derived from the following equation by the maximum-limit force of fracture [37]:

$$K_{IC} = \frac{3FL\sqrt{c \cdot 10^{-3}}}{2BW^2}\left[1.93 - 3.07(\frac{c}{W}) + 14.53(\frac{c}{W})^2 - 25.07(\frac{c}{W})^3 + 25.80(\frac{c}{W})^4\right] \quad (4)$$

Where *F*, *L*, *B*, *W*, and c are the breaking load, supporting span, width, height, and the pre-crack length, respectively. Three test pieces were tested to get the average value for each composition.

The room-temperature Young's modulus (E) was measured using a commercial resonant ultrasound spectroscopy (RUS) apparatus with $E = \frac{\rho V_s^2 (3T^2 - 4)}{T^2 - 1}$, where $V_L$ is longitudinal wave velocity, $V_S$ is shear wave velocity and $T$ is their ratio, i.e. $T = V_L/V_S$. Vickers indentation was conducted at a load of force (20 N) using a measuring system (VMHK-30, Japan).

**First-principles calculation method**

First-principles calculations are performed by VASP 5.4 with a cutoff energy of 520 eV and a K point mesh of 7×7×7. The vacancy was implemented in a Mn doped 3×3×2 $Mg_3Sb_2$ supercell with the Mn ion occupying an interstitial site below an Mg2 site, which is the most energetic favorable doping configuration. The formation energies were calculated based on the supercell method [54]. More details can be found in ref. [55]. Various charge states of vacancies are considered, the formation energy of -2 charge states are found to be the lowest in $E_F$ in experiment range.

**Aknowledgement**

The authors would like to thank the support of Pearl River Talents Recruitment Program No.2016ZT06G587, National 1000-Talent Plan and Shenzhen Peacock Talent Plan. We are also grateful to Prof. Jiaqing He for providing great help in sample preparation and measurement about mechnical properties.

## Caption of Figures:

**Figure 1** (a-b) Schematic figure of the band edge structures of $Mg_3Sb_{0.5}Bi_{1.5}$ and $Mg_3Sb_2$, which shows the resolution of $CB_1$ band (0, 0.417, 0.333) according to the ref. [24-25]. (c) Optical band gaps of as-fabricated $Mg_{3+\delta}Sb_xBi_{2-x}$ family, compared the theoretical calculated values. (d) the $(\alpha h\nu)^2$ versus $h\nu$ plot of samples $Mg_{3+\delta}Sb_xBi_{2-x}$ (x=1.5, x=1.3 and 1.1) at room temperature.

**Figure 2** (a) Mg1 and Mg2 vacancy formation energies in intrinsic and Mn doped crystals under different conditions. (b) Experimental (Exp.) and theoretically fitting Mn K-edge absorption spectra for five possible Mn positions, including interstitial sites: IA (2/3, 1/3, 0.182; or 1/3, 2/3, 0.581) and IC (0, 1/2, 1/2); substitution-sites: Mg1(0, 0, 0), Mg2 (1/3, 2/3, 0.631; or 2/3, 1/3, 0.368) and Sb (1/3, 2/3, 0.225; or 2/3 1/3 0.775). (c) Crystal structure of intrinsic $Mg_{3+\delta}Sb_xBi_{2-x}$ and $Mg_{3+\delta}Sb_xBi_{2-x}$: $Mn_{0.01}$.

**Figure 3** (a-c) Thermoelectric properties of the as-fabricated classical composition $Mg_{3+\delta}Sb_xBi_{1.99-x}Te_{0.01}$ (x=1.5 and x=2) compared with data from ref. [21, 28-29]: (a) electrical resistivity, (b) Seebeck coefficient, (c) Power factor. (d-f) Thermoelectric transport properties of $Mg_{3+\delta}Sb_xBi_{1.99-x}Te_{0.01}$: $Mn_{0.01}$ (x=1.1-1.5) in directions perpendicular (⊥) to press direction, (d) electrical resistivity, (e) Seebeck coefficient, (f) power factor.

**Figure 4** Temperature dependent of (a) thermal conductivity κ and (b) figure of merit *ZT* of as-fabricated $Mg_{3+\delta}Sb_xBi_{1.99-x}Te_{0.01}$: $Mn_{0.01}$ (x=1.1-1.5) in directions perpendicular (⊥) to press direction, additionally, the thermal conductivity and electrical conductivity in directions parallel (//) to press direction were given in Figure S2, S8 respectively. (c) is the temperature dependent band gap for the samples $Mg_{3+\delta}Sb_xBi_{1.99-x}Te_{0.01}$: $Mn_{0.01}$ (x=1.1-1.5). (d) Temperature dependence of band gap $dE_g/dT$ for as-fabricated $Mg_{3+\delta}Sb_xBi_{1.99-x}Te_{0.01}$: $Mn_{0.01}$ (x=1.1-1.5) and selected semiconductors (the detailed data and references were given in the Supporting Information).

**Figure 5** Thermoelectric and mechanical properties of as-fabricated $Mg_{3+\delta}Sb_xBi_{2-x}$ family. (a) Temperature dependent figure of merit, (b) engineering figure of merit in the temperature range of 50-250 °C, (c) average figure of merit in the temperature range of 50-250 °C.

**Figure 6** (a) Fracture toughness against Young's modulus of typical thermoelectric materials. (b) Composition dependent fracture toughness. The inset displays indentation crack patterns on the bulk surfaces of $Mg_{3+\delta}Sb_{1.2}Bi_{0.8}$: $Mn_{0.01}$formed under the load of 20 N, in which the cracks flaked off near the edge of indentation partly. The detailed reference data are given in main text and support information.

**Figure 7** Microstructure of as-fabricated $Mg_{3+\delta}Sb_{x}Bi_{1.99-x}Te_{0.01}$: $Mn_{0.01}$. (a) typical SEM image of fresh fractured surface. (b, c) Low-magnification TEM image of the sample with Sb content of x=1.2. Inset shows the electron diffraction patterns along $[1\bar{2}1\bar{3}]$. (d) High angle annular dark field (HAADF) of the coarse-grained boundary. The enlarged box view in (d) shows the FFT image of the corresponding boxed region.

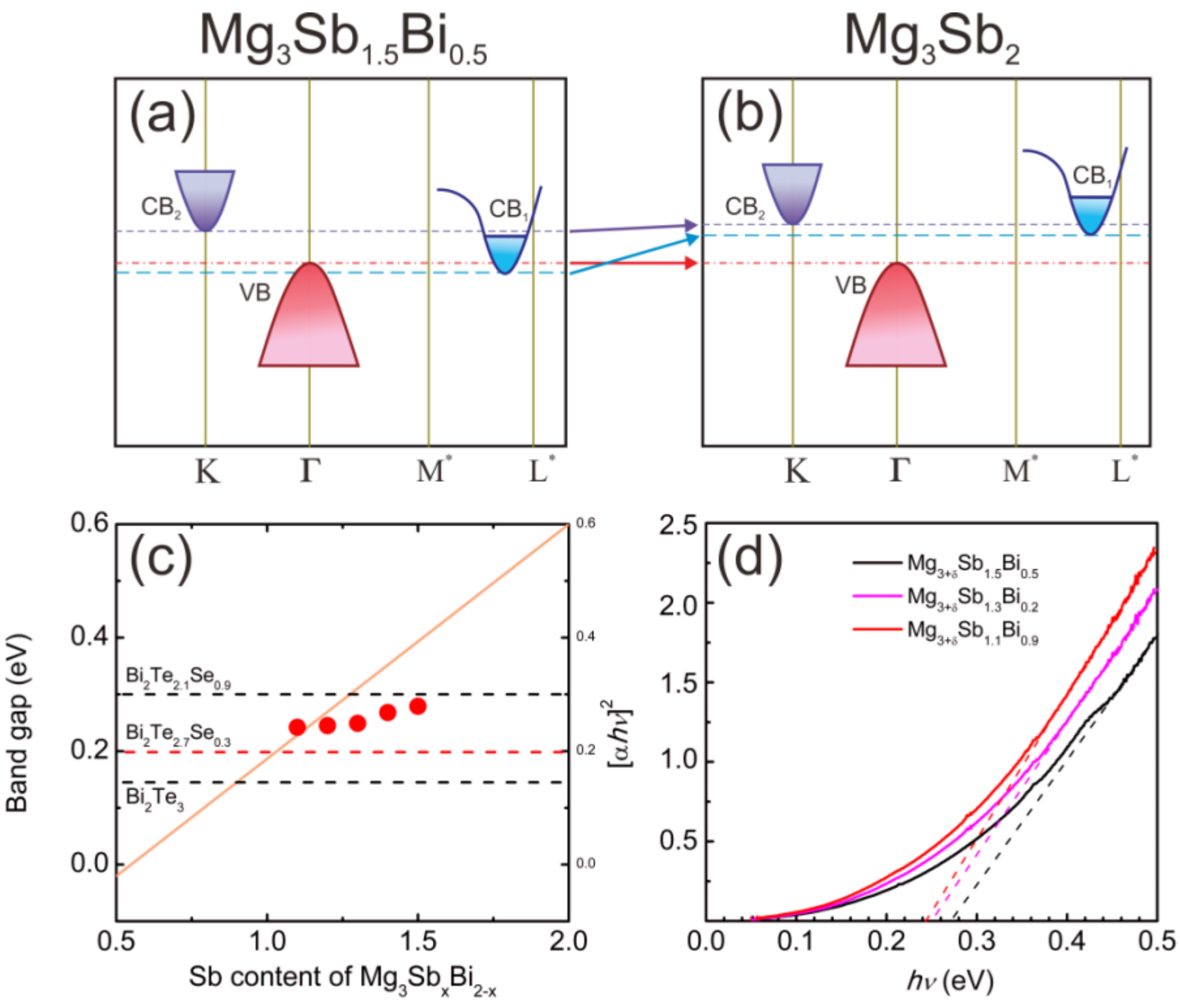

**Figure 1**

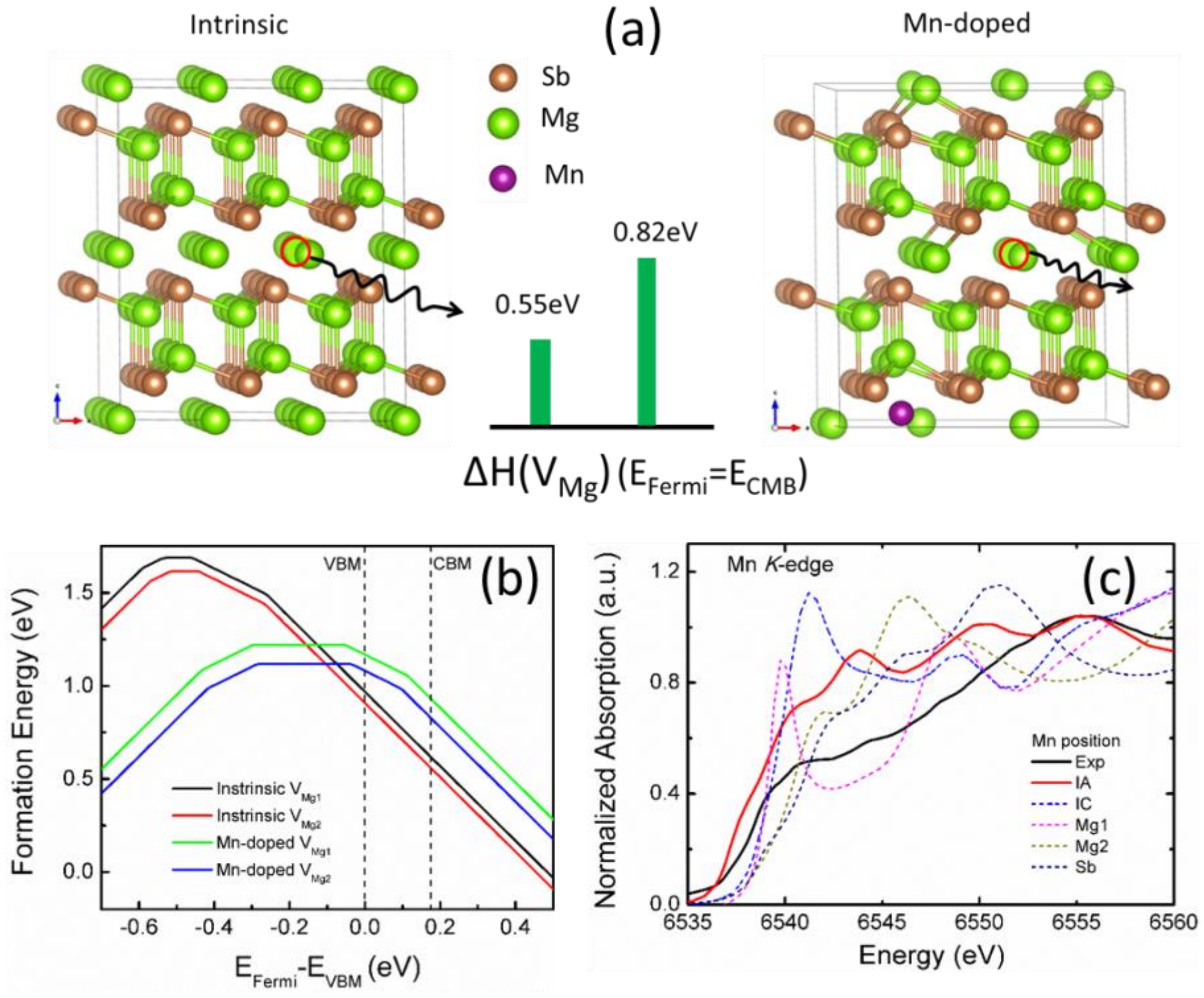

**Figure 2**

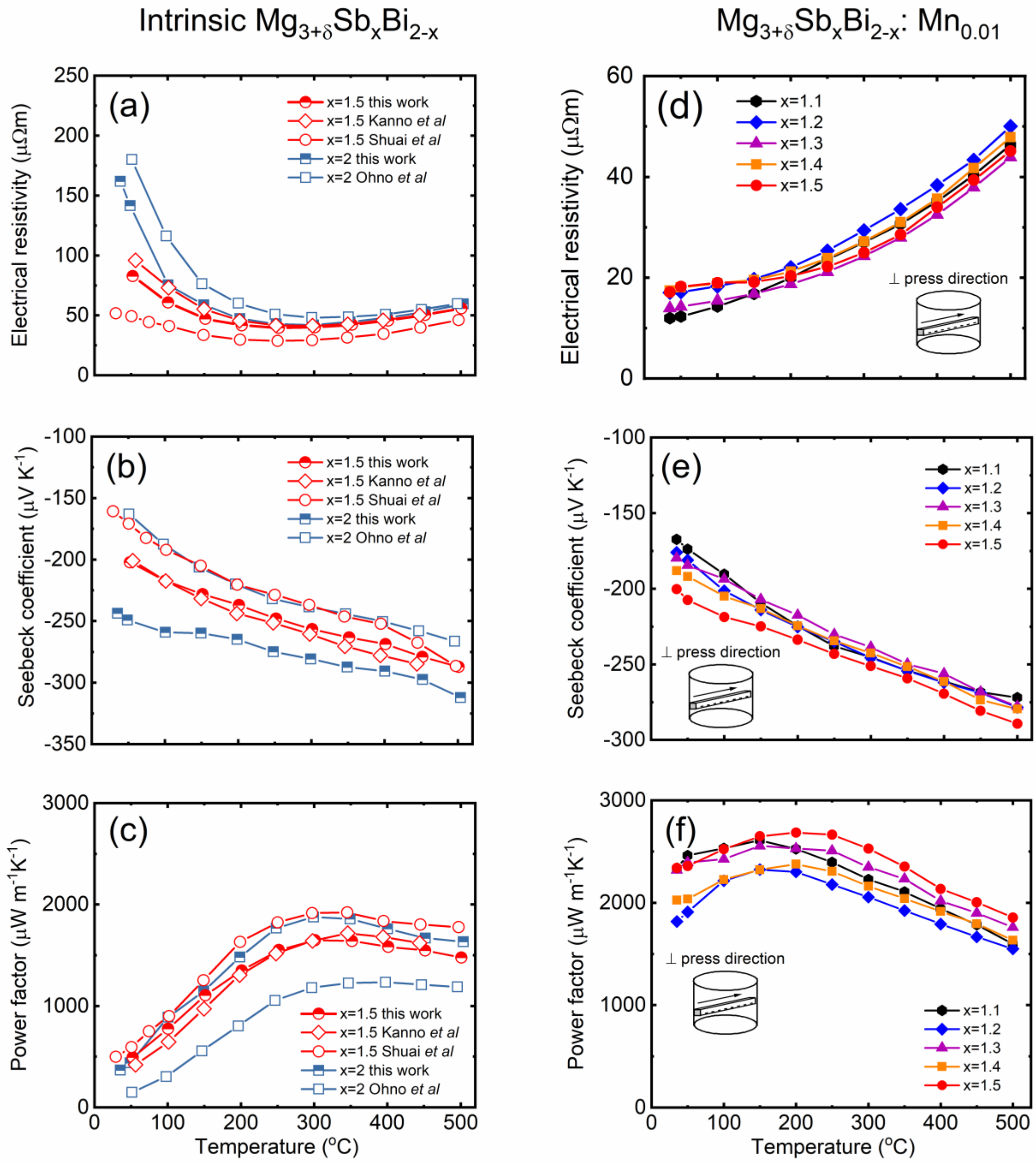

Intrinsic Mg3+δSbxBi2-x
Mg3+δSbxBi2-x: Mn0.01
(a)
x=1.5 this work
x=1.5 Kanno et al
x=1.5 Shuai et al
x=2 this work
x=2 Ohno et al
Electrical resistivity (μΩm)
(d)
x=1.1
x=1.2
x=1.3
x=1.4
x=1.5
⊥ press direction
(b)
Seebeck coefficient (μV K-1)
(e)
(c)
Power factor (μW m-1K-1)
(f)
Temperature (°C)

**Figure 3**

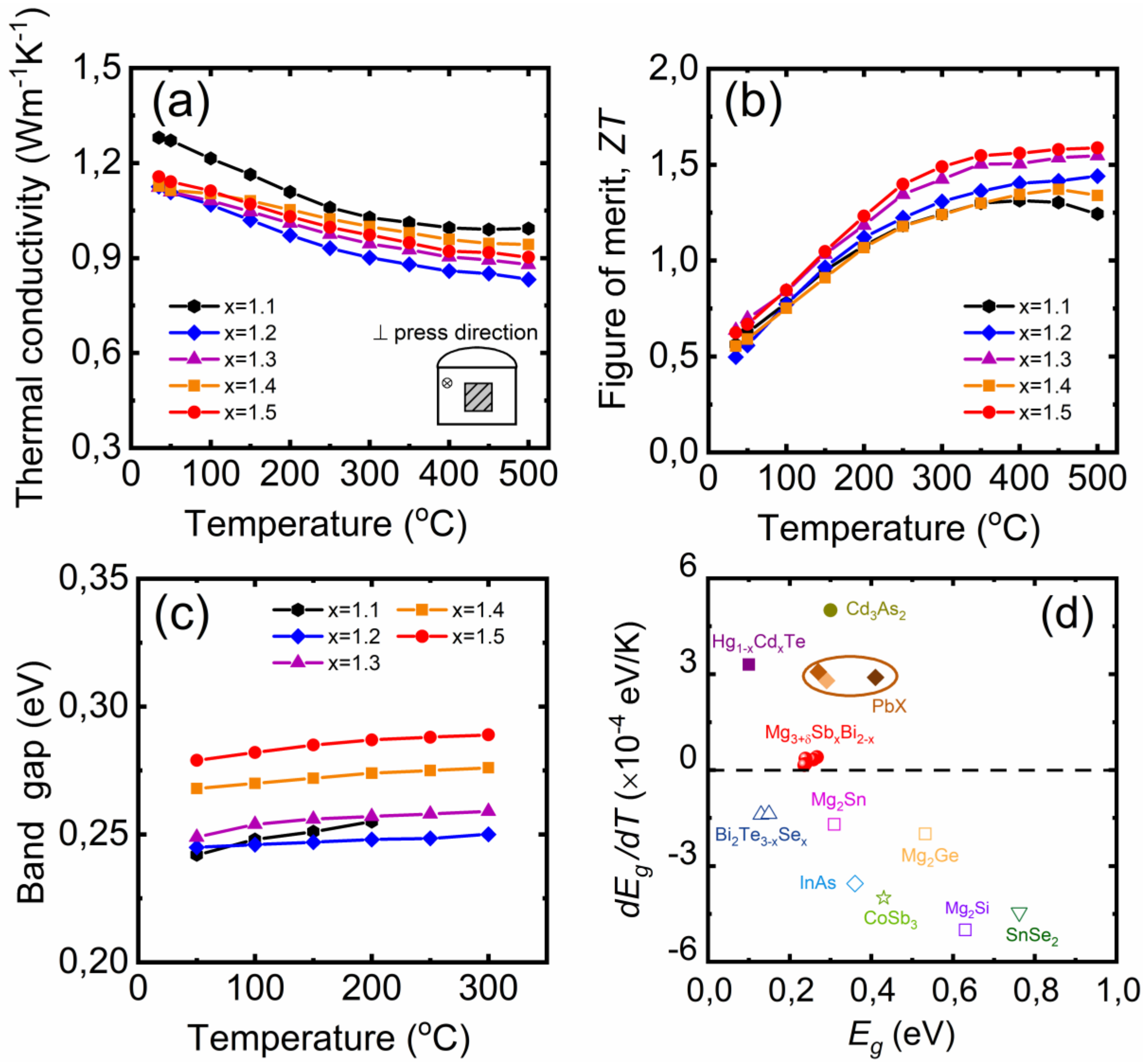

**Figure 4**

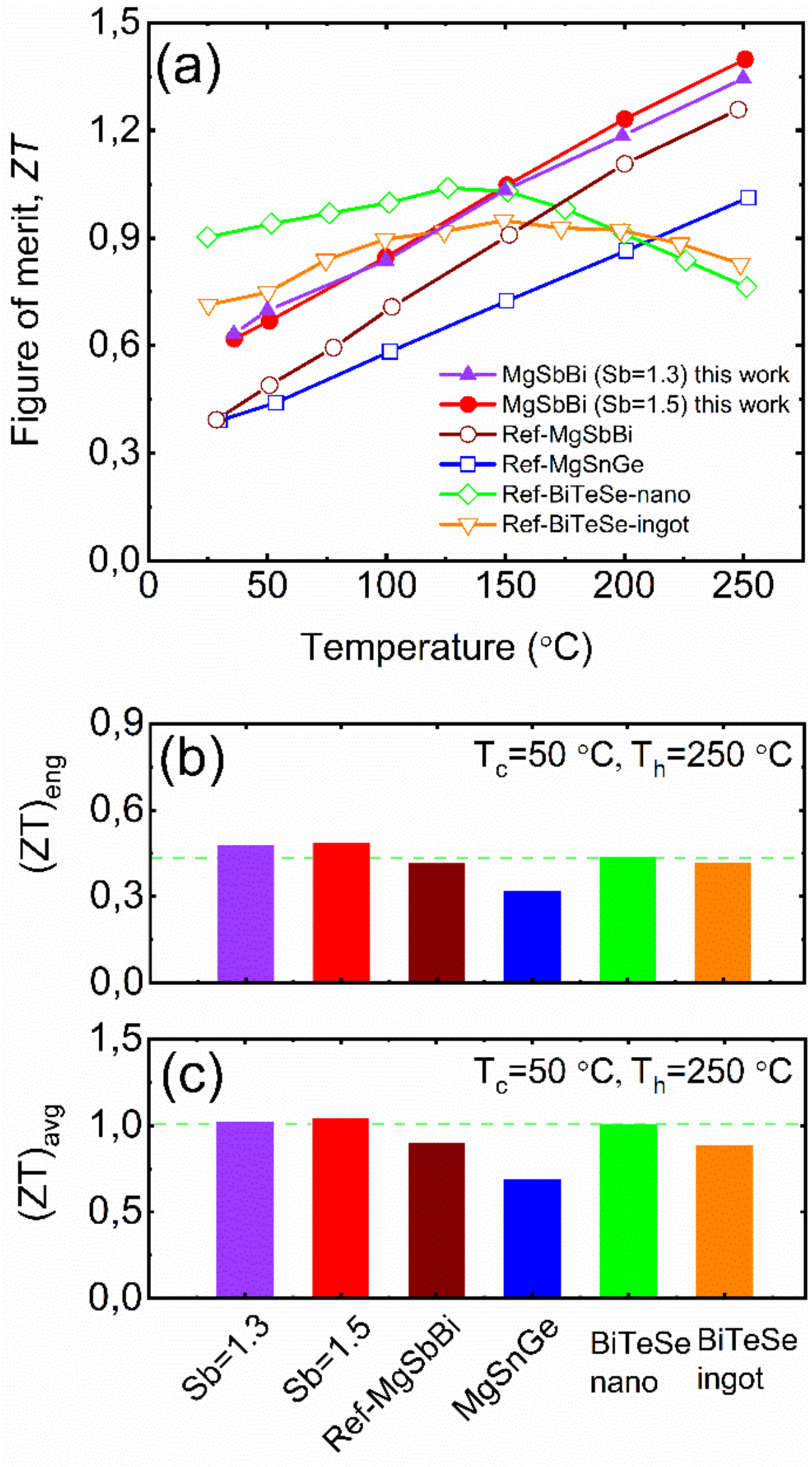

**Figure 5**

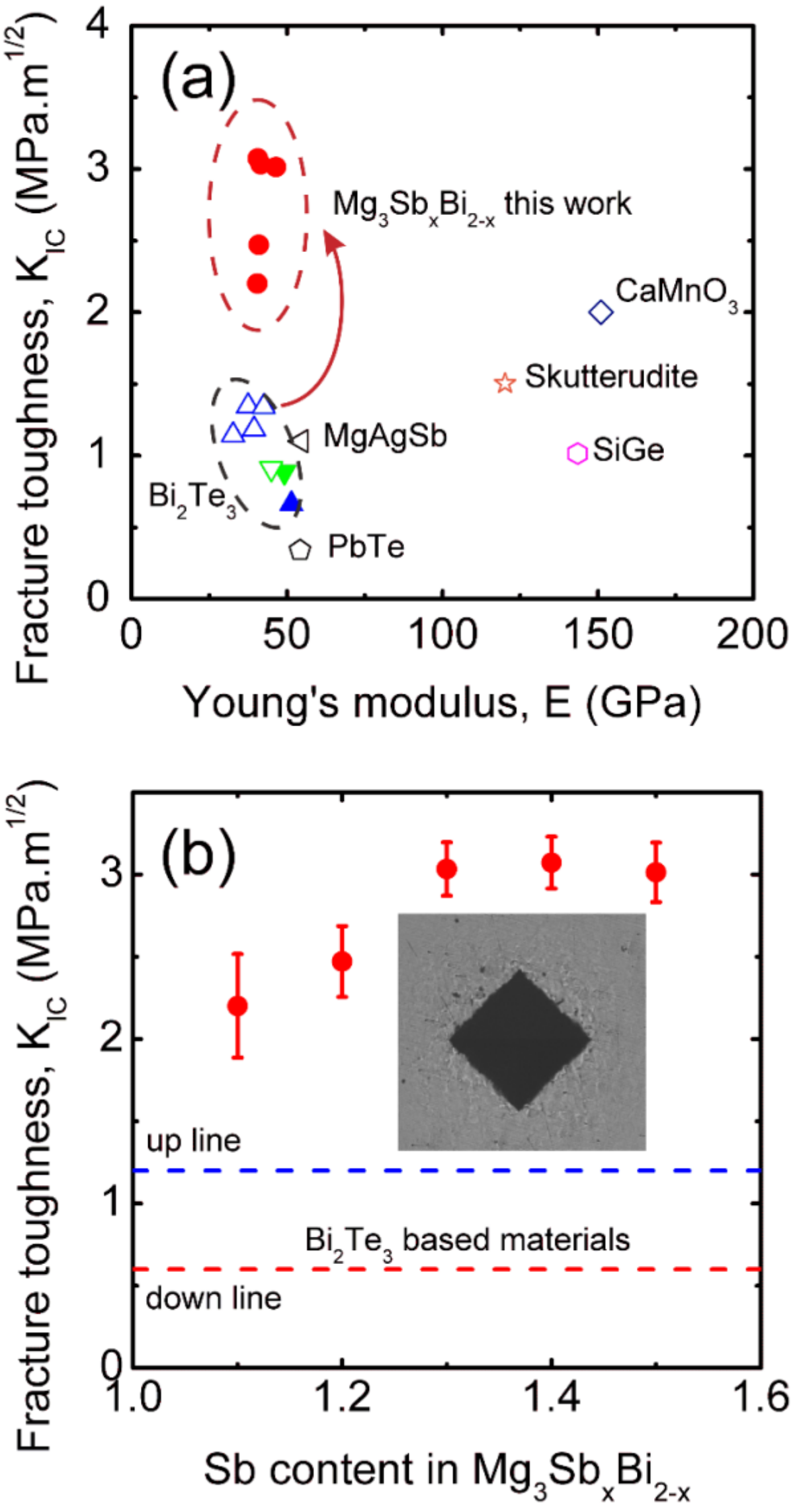
(a)
Fracture toughness, $K_{IC}$ ($MPa.m^{1/2}$)
Young's modulus, E (GPa)
$Mg_3Sb_xBi_{2-x}$ this work
$CaMnO_3$
Skutterudite
MgAgSb
SiGe
$Bi_2Te_3$
PbTe
(b)
Fracture toughness, $K_{IC}$ ($MPa.m^{1/2}$)
Sb content in $Mg_3Sb_xBi_{2-x}$
up line
$Bi_2Te_3$ based materials
down line

**Figure 6**

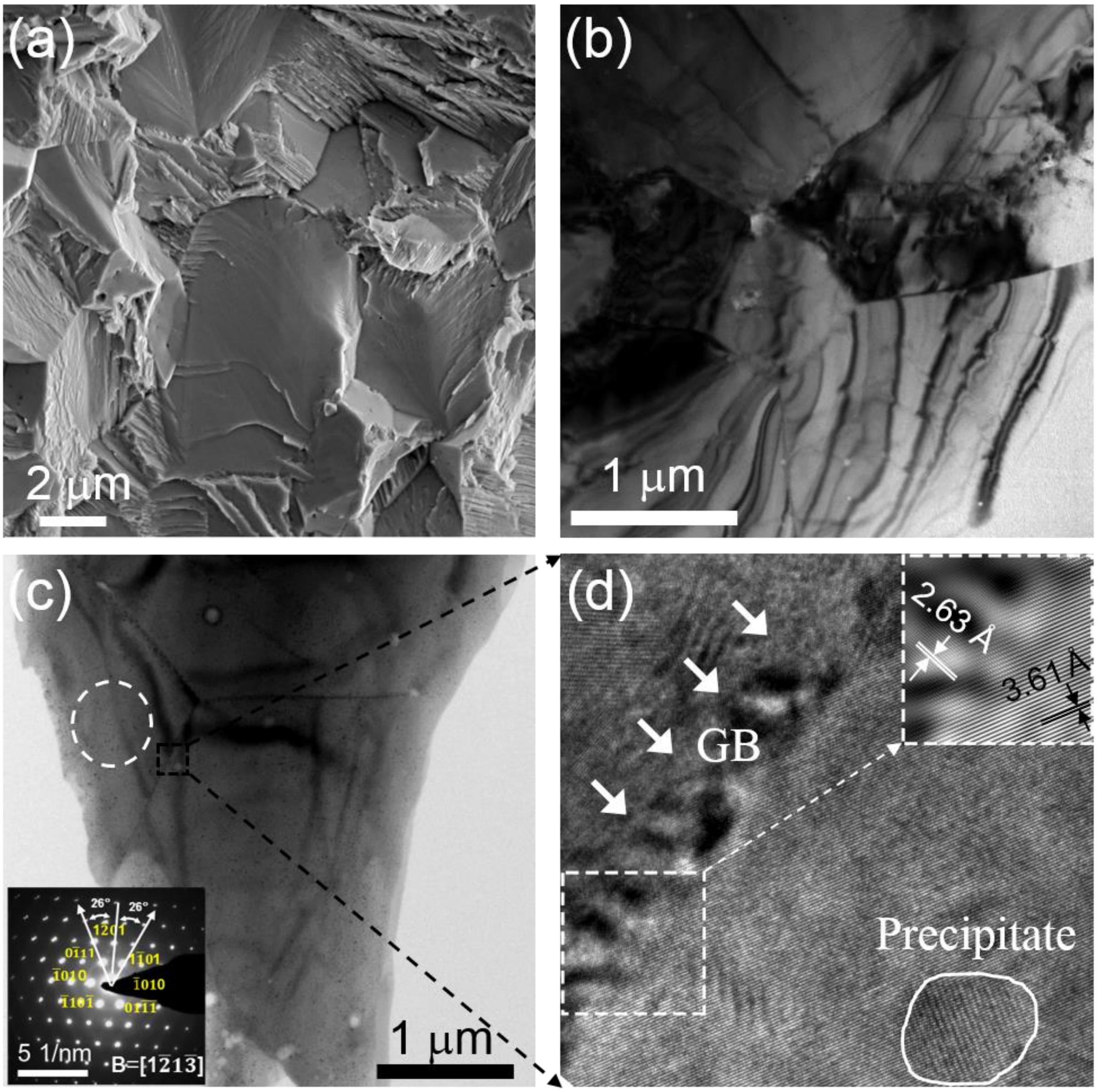

**Figure 7**

*Supporting Information for:*

# $Mg_{3+\delta}Sb_xBi_{2-x}$ family: A promising substitute for the start-of-art n-type thermoelectric materials near room temperatures

Rui Shu[1], Yecheng Zhou[2], Qi Wang[1], Zhijia Han[1], Yongbin Zhu[1], Yong Liu[1,3], Yuexing Chen[2,6], Meng Gu[1], Wei Xu[4], Yu Wang[5], Wenqing Zhang[2], Li Huang[2*] and Weishu Liu[1*]

[1]Department of Materials Science and Engineering, Southern University of Science and Technology, Shenzhen 518055, China

[2]Department of Physics, Southern University of Science and Technology, Shenzhen 518055, China

[3]Beijing Institute of Aeronautical Materials, AECC, Beijing 100095, China

[4]Beijing Synchrotron Radiation Facility, Institute of High Energy Physics, Chinese Academy of Sciences, Beijing 100049, China

[5]Shanghai Synchrotron Radiation Facility, Shanghai Institute of Applied Physics, Chinese Academy of Sciences, Shanghai 201204, China

[6]Shenzhen Key Laboratory of Advanced Thin Films and Applications, College of Physics and Energy, Shenzhen University, Shenzhen 518060, China

*E-mail: huangl@sustc.edu.cn, liuws@sustc.edu.cn

**Supplementary figures:**

**Figure S1** (a) X-ray diffraction (XRD) patterns of as-fabricated $Mg_{3+\delta}Sb_xBi_{2-x}$ family. (b) Lattice parameters of as-fabricated $Mg_{3+\delta}Sb_xBi_{2-x}$ family as a function of the fraction x. The nominal composition is: $Mg_{3+\delta}Sb_xBi_{1.99-x}Te_{0.01}$: $Mn_{0.01}$ (x=1.1, 1.2, 1.3, 1.4 and 1.5). No notable impurity phase, such as excess Mg, was identified.

**Figure S2 (a)** Temperature dependence of thermal diffusivity for n-type $Mg_{3+\delta}Sb_xBi_{1.99-x}Te_{0.01}$(x=1.1, 1.2, 1.3, 1.4 and 1.5) in directions perpendicular to press direction and $Mg_{3+\delta}Sb_xBi_{1.99-x}Te_{0.01}$(x=1.2, 1.5) in parallel direction to the press. (b) Temperature dependence of specific heat for n-type $Mg_{3+\delta}Sb_xBi_{1.99-x}Te_{0.01}$(x=1.1, 1.2, 1.3, 1.4 and 1.5). The line represents the value calculated from the Dulong-Petit formula of $Mg_{3+\delta}Sb_{1.5}Bi_{0.49}Te_{0.01}$.

**Figure S3** The absorption spectra of $Mg_{3+\delta}Sb_{1.2}Bi_{0.8}$, plotted as a function of the energy *hv* in the range of 50 °C to 300 °C. The across point of the dash line with *hv* axis was used to determine the band gap. The inset presents the slight move of spectra curve with the increasing of temperature.

**Figure S4** The reproducibility of thermoelectric properties of $Mg_{3.2}Sb_{1.5}Bi_{0.49}Te_{0.01}$: $Mn_{0.01}$ in ⊥ direction to press direction (three batch samples, including two cut from 3 mm thickness disks and one cut from 10 mm thickness disk): (a) electrical resistivity; (b) Seebeck coefficient; (c) power factor; (d) thermal diffusion.

**Figure S5** The reproducibility of thermoelectric properties of $Mg_{3.2}Sb_{1.2}Bi_{0.79}Te_{0.01}$: $Mn_{0.01}$ (three batch samples, including two cut from 3 mm thickness disks and one cut from 10 mm thickness disk). (a) electrical resistivity; (b) Seebeck coefficient; (c) power factor; (d) thermal diffusion.

**Figure S6** The heating and cooling measurements for n-type $Mg_{3.2}Sb_{1.2}Bi_{0.79}Te_{0.01}$: $Mn_{0.01}$. (a) electrical conductivity; (b) Seebeck coefficient; (c) power factor.

**Figure S7** Comparison of thermoelectric properties between as-fabricated $Mg_{3+\delta}Sb_xBi_{2-x}$ compounds (x=1.2 and 1.5) and selected n-type materials. Ref.-MgSbBi: Nb doped $Mg_{3.2}Sb_{1.5}Bi_{0.5}$[1], Ref.-MgSnGe: $Mg_2Sn_{0.75}Ge_{0.25}$[2], Ref.-BiTeSe-nano: $Bi_2Te_{0.79}Se_{0.21}$[3], Ref.-BiTeSe-ingot, ordered from a company in China.

**Figure S8** Thermoelectric properties of $Mg_{3.2}Sb_xBi_{1.99-x}Te_{0.01}$: $Mn_{0.01}$ (x=1.2, 1.5) in directions perpendicular (⊥) to press direction and directions parallel (//) to press direction: (a) electrical resistivity; (b) Seebeck coefficient; (c) thermal diffusion; (d) thermal conductivity; (e) power factor; (f) figure of merit.

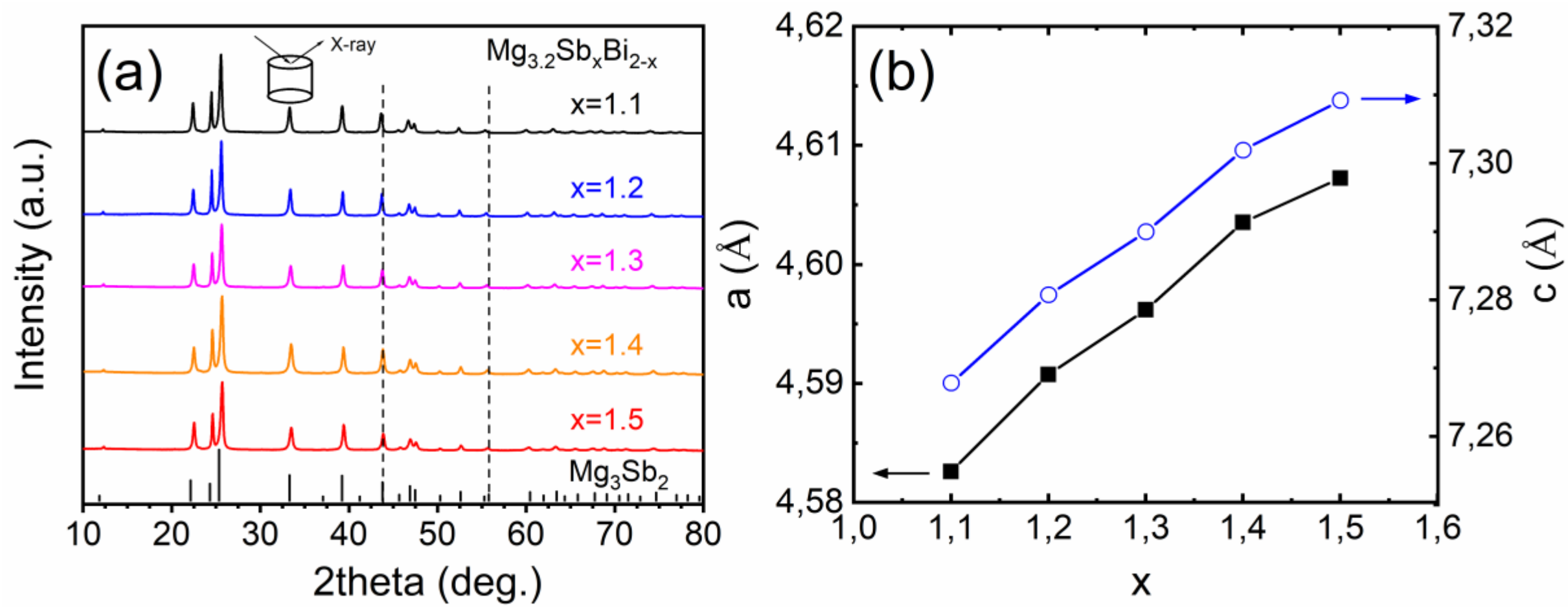

(a)
X-ray
Mg3.2SbxBi2-x
x=1.1
x=1.2
x=1.3
x=1.4
x=1.5
Mg3Sb2
Intensity (a.u.)
2theta (deg.)
10 20 30 40 50 60 70 80
(b)
a (Å)
c (Å)
x
4,58 4,59 4,60 4,61 4,62
7,26 7,28 7,30 7,32
1,0 1,1 1,2 1,3 1,4 1,5 1,6

**Figure S1**

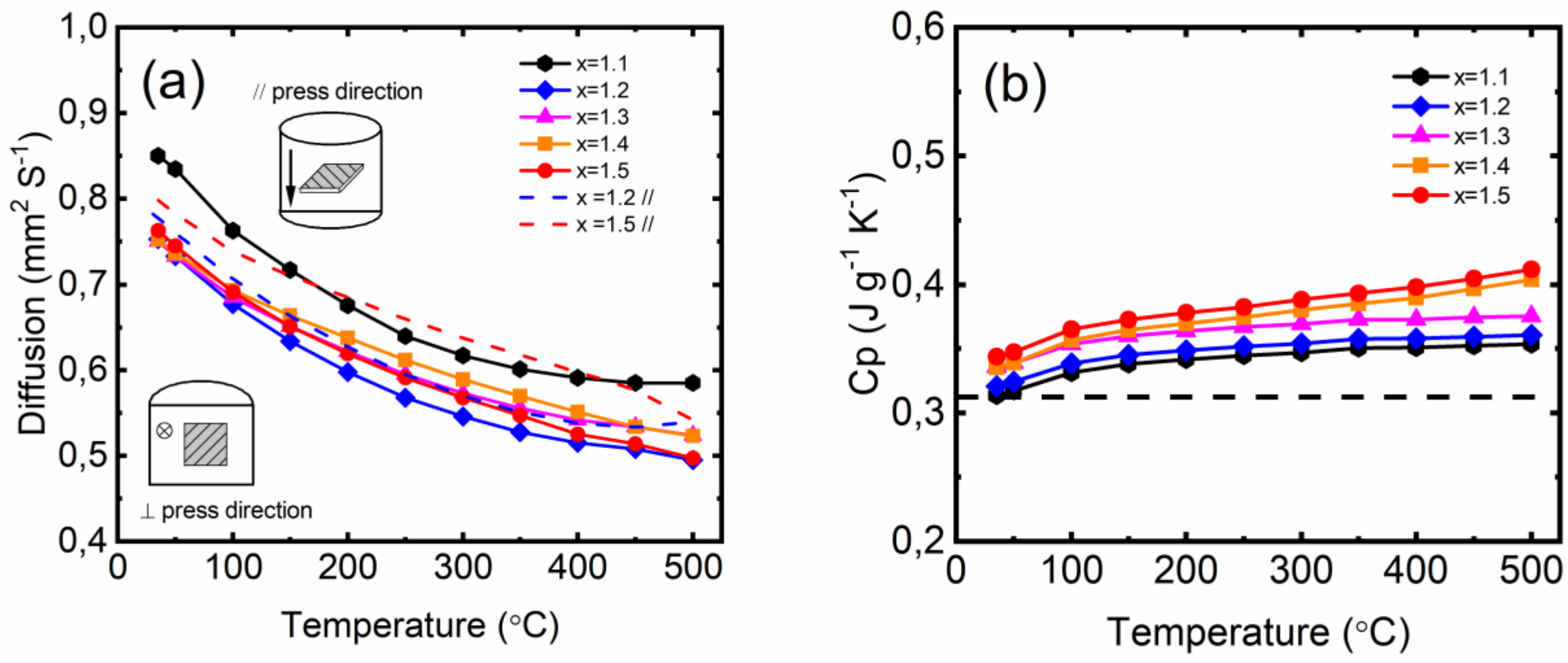

**Figure S2**

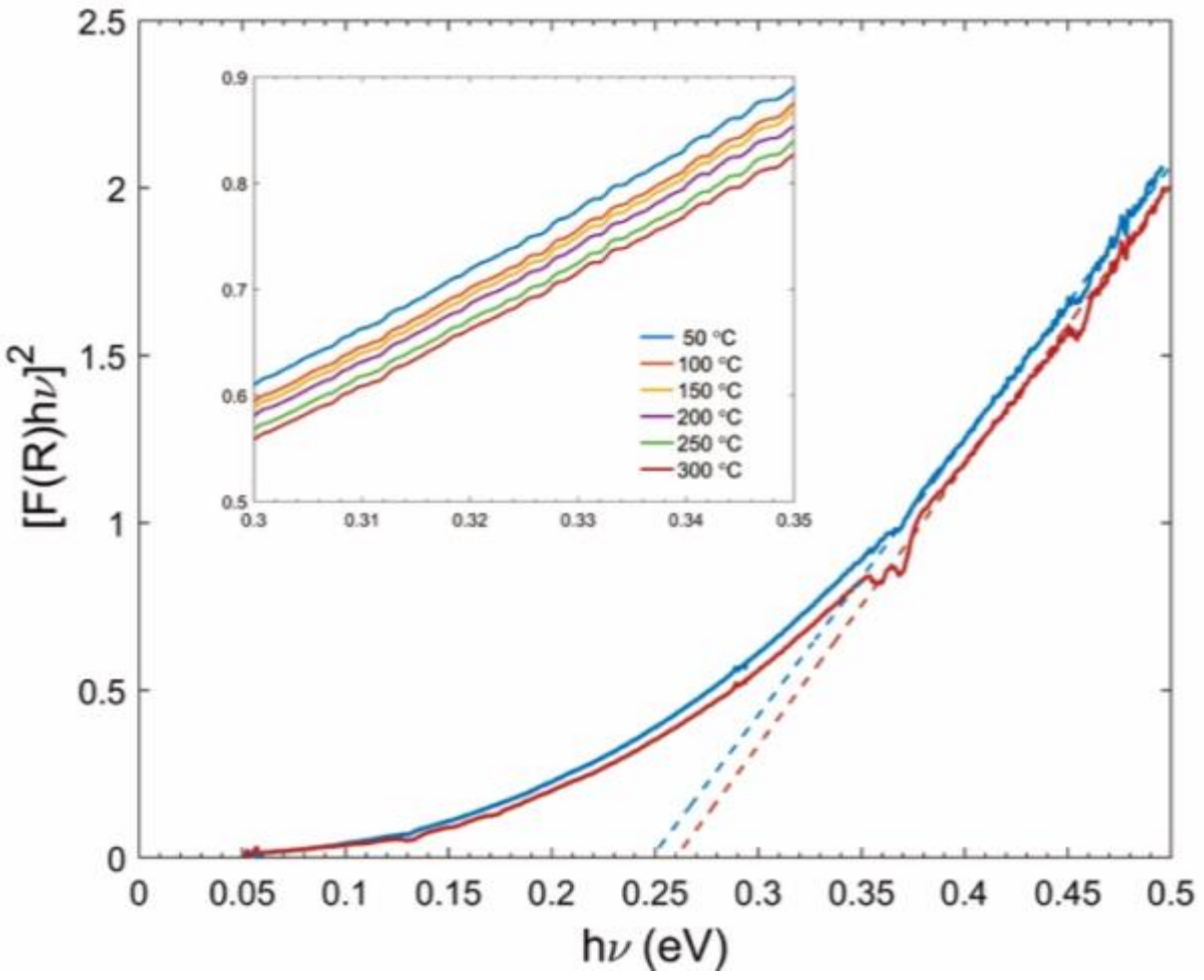

**Figure S3**

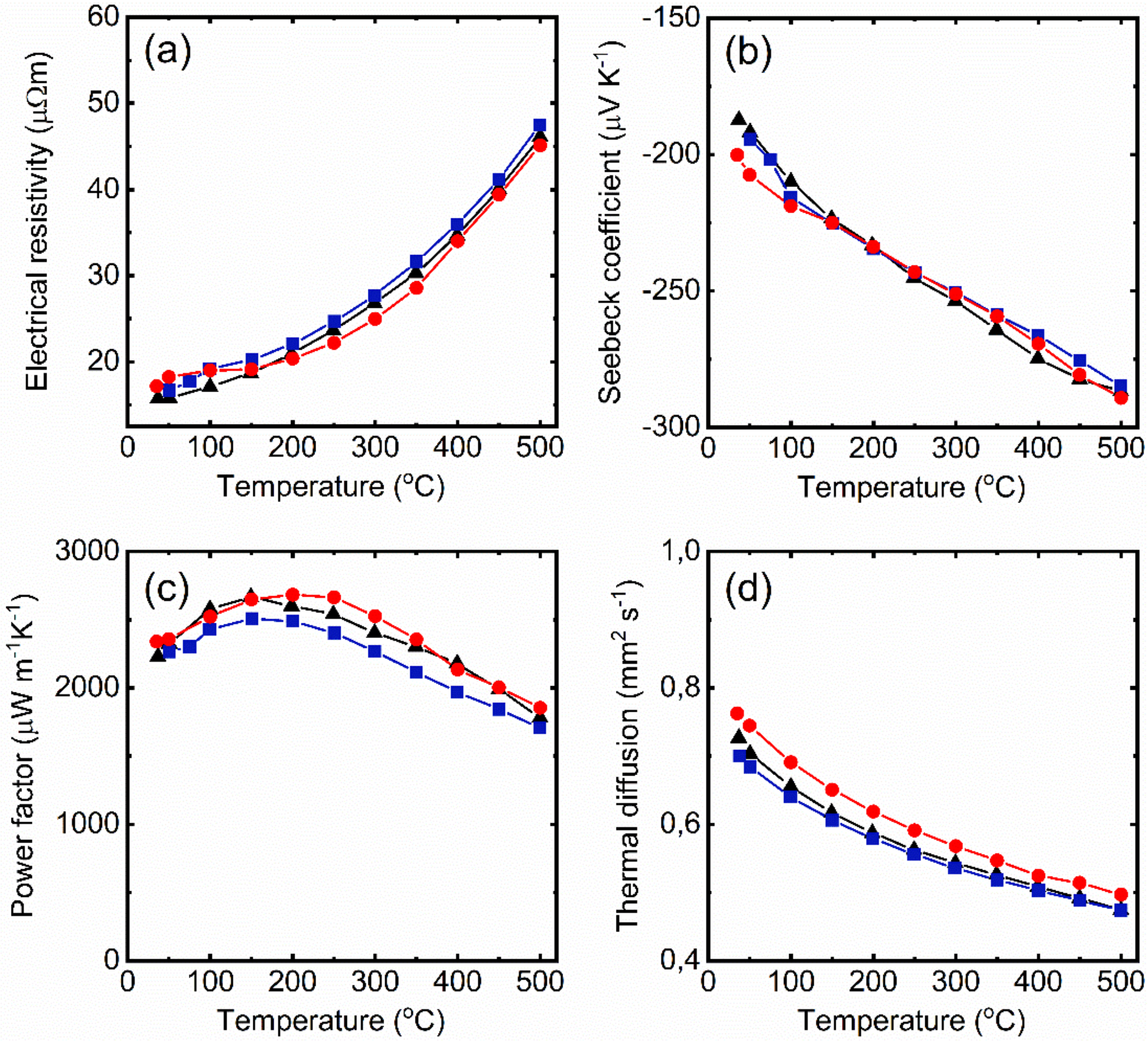
(a)
Electrical resistivity (μΩm)
Temperature (°C)
(b)
Seebeck coefficient (μV K-1)
Temperature (°C)
(c)
Power factor (μW m-1K-1)
Temperature (°C)
(d)
Thermal diffusion (mm2 s-1)
Temperature (°C)

**Figure S4**

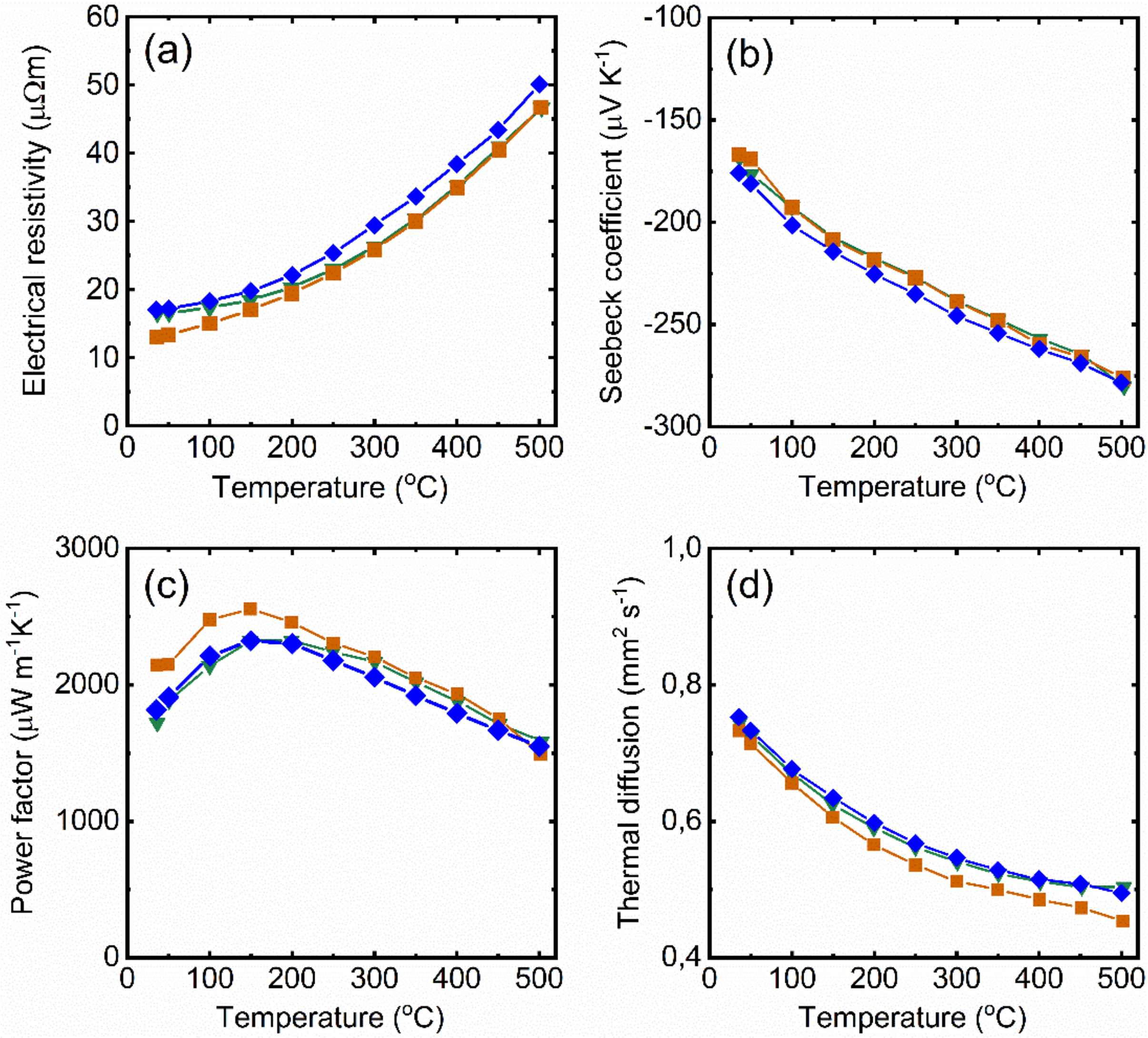

(a)
Electrical resistivity (μΩm)
Temperature (°C)
(b)
Seebeck coefficient (μV K⁻¹)
Temperature (°C)
(c)
Power factor (μW m⁻¹K⁻¹)
Temperature (°C)
(d)
Thermal diffusion (mm² s⁻¹)
Temperature (°C)

**Figure S5**

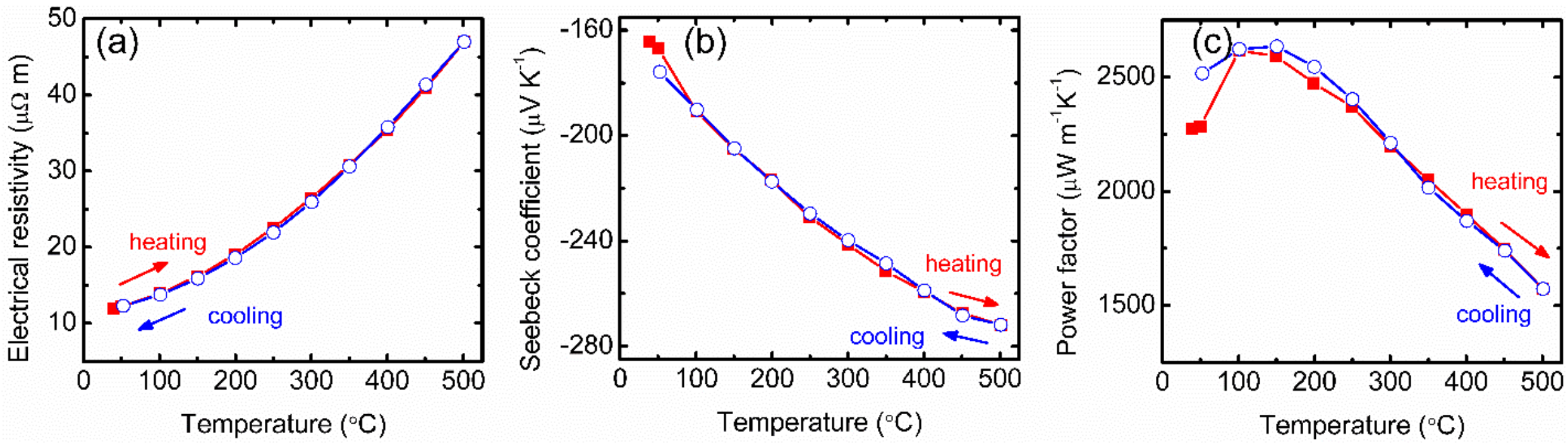

**Figure S6**

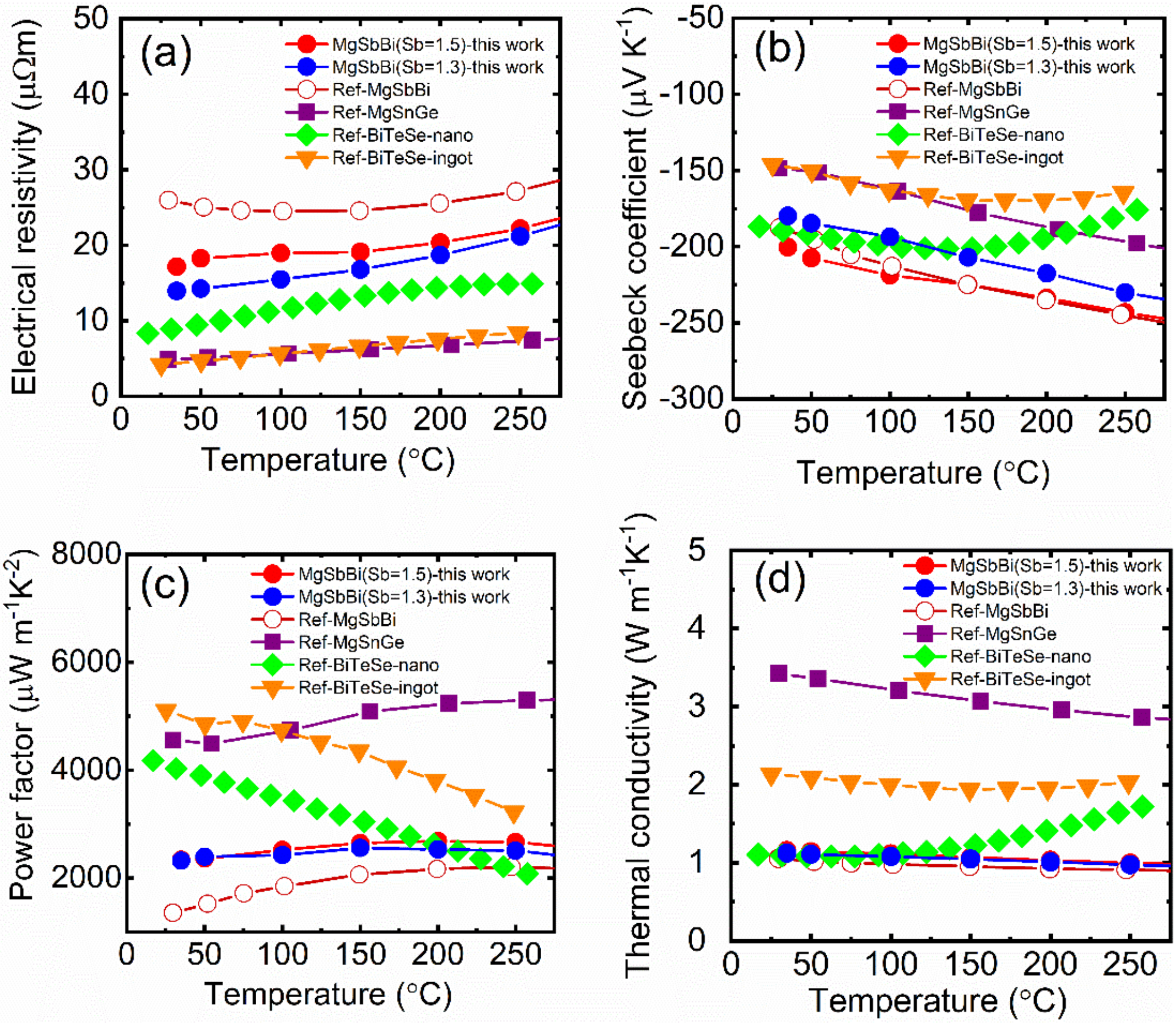

**Figure S7**

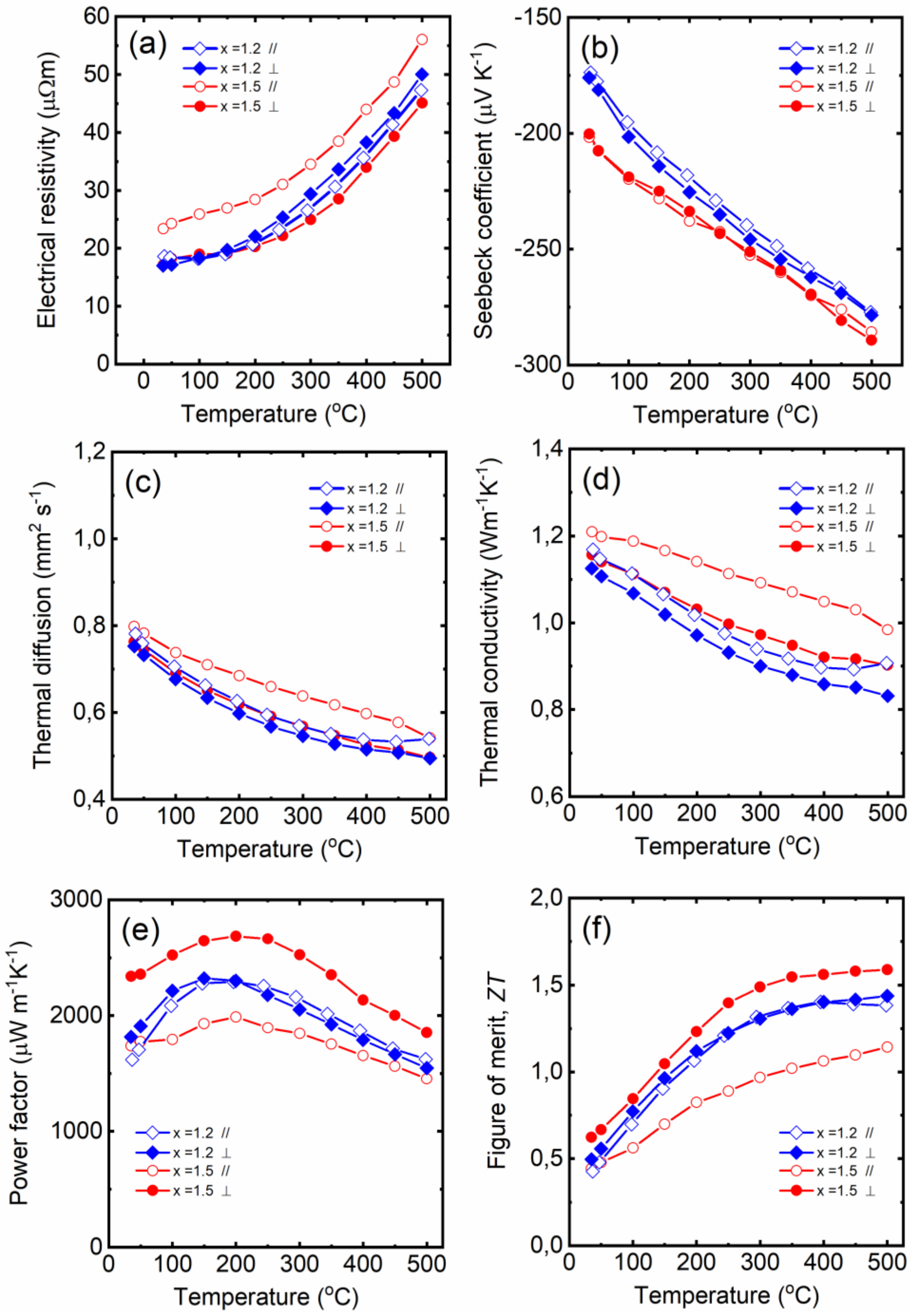
(a)
Electrical resistivity (μΩm)
Temperature (°C)
x =1.2 //
x =1.2 ⊥
x =1.5 //
x =1.5 ⊥
(b)
Seebeck coefficient (μV K-1)
(c)
Thermal diffusion (mm2 s-1)
(d)
Thermal conductivity (Wm-1K-1)
(e)
Power factor (μW m-1K-1)
(f)
Figure of merit, ZT

**Figure S8**

**Table S1.** The temperature dependence of band gap ($E_g=a+b\times T$), $E_0$ (band gap at RT) and slope $b$ valves for typical semiconductor materials.

| ***Materials*** | ***$E_0$ (eV)*** | ***b ($10^{-4}$ eV/K)*** | ***Ref.*** |
|---|---|---|---|
| $Hg_{1-x}Cd_xTe$ | 0.100 | 3.30 | [4] |
| PbTe | 0.291 | 2.82 | [5] |
| PbSe | 0.270 | 3.07 | |
| PbS | 0.408 | 2.90 | |
| $Mg_{3.2}Sb_{1.5}Bi_{0.6}$ | 0.267 | 0.40 | *this work* |
| $Mg_{3.2}Sb_{1.4}Bi_{0.7}$ | 0.258 | 0.33 | |
| $Mg_{3.2}Sb_{1.3}Bi_{0.7}$ | 0.239 | 0.36 | |
| $Mg_{3.2}Sb_{1.2}Bi_{0.8}$ | 0.2385 | 0.19 | |
| $Mg_{3.2}Sb_{1.1}Bi_{0.9}$ | 0.236 | 0.15 | |
| $Bi_2Te_3$ | 0.130 | -1.40 | [6] |
| $Bi_2Te_{2.7}Se_{0.3}$ | 0.149 | -1.37 | [6] |
| $Mg_2Sn$ | 0.309 | -1.70 | [6] |
| $Mg_2Ge$ | 0.532 | -2.00 | |
| $Mg_2Si$ | 0.630 | -5.00 | |
| $SnSe_2$ | 0.762 | -4.45 | [7] |
| $CoSb_3$ | 0.43 | -4.00 | [8] |
| InAs | 0.36 | -3.55 | [6] |

**Table S2.** The materials cost of $Mg_3Sb_xBi_{2-x}$ (x=1.1, 1.5) and several common n-type thermoelectric materials.

| *n-type TE materials* | *Materials Cost(($/kg))* |
|---|---|
| $Bi_2Te_{2.7}Se_{0.3}$ | 44.1 |
| PbTe | 30.6 |
| $CoSb_3$ | 10.3 |
| SiGe | 1805.2 |
| $Mg_2Si_{0.85}Bi_{0.15}$ | 7.9 |
| $Mg_2Sn_{0.75}Ge_{0.25}$ | 157.3 |
| **$Mg_3Sb_{1.1}Bi_{0.9}$** | **11.3** |
| **$Mg_3Sb_{1.5}Bi_{0.5}$** | **9.5** |
| Mg | 4.74 |
| Bi | 16.7 |
| Te | 77 |
| Se | 48.7 |
| Co | 29.2 |
| Sb | 7.2 |
| Si | 2.52 |
| Sn | 16.7 |
| Pb | 2.01 |
| Ge | 1250 |

**Table S3**. The specific heat of the $Mg_{3.2}Sb_xBi_{1.99-x}Te_{0.01}$: $Mn_{0.01}$ (x=1.1-1.5)

| ***Sample*** | $C_p=AT^3+BT^2+CT+D$ | | | |
|---|---|---|---|---|
| | ***A*** | ***B*** | ***C*** | ***D*** |
| x=1.5 | -4.8702E-11 | -2.3606E-07 | 2.4803E-04 | 3.0208 |
| x=1.4 | -4.8702E-11 | -2.3606E-07 | 2.4803E-04 | 3.0126 |
| x=1.3 | -4.8702E-11 | -2.3606E-07 | 2.4803E-04 | 3.0047 |
| x=1.2 | -4.8702E-11 | -2.3606E-07 | 2.4803E-04 | 2.9972 |
| x=1.1 | -4.8702E-11 | -2.3606E-07 | 2.4803E-04 | 2.9901 |

**Supplementary Note I. Relationship between $(ZT)_{avg}$, $(ZT)_{eng}$ and leg efficiency**

The thermoelectric figure of merit $ZT$ is defined through the derivation without considering temperature dependent transport properties, which is a simplified scale for the devices level performance, i.e. leg efficiency [9].

$$\eta=\frac{T_h-T_c}{T_h}\frac{\sqrt{Z\bar{T}+1}-1}{\sqrt{Z\bar{T}+1}+\frac{T_c}{T_h}} \quad \text{(S1)}$$

$T_h$ and $T_c$ are the hot side temperature and cold side temperature, respectively. $\bar{T}$ is the average temperature between the $T_h$ and $T_c$.

In real case, all the thermoelectric parameters, including Seebeck coefficient, electrical conductivity and thermal conductivity, show a significantly temperature dependence. The thermoelectric performance evaluation is therefore necessary considered the temperature dependence by using the average figure-of-merit, i.e. $(ZT)_{avg}$ within the considered temperature range as following:

$$(ZT)_{avg}=\frac{1}{T_h-T_c}\int_{T_c}^{T_h} ZT(T)dT \quad \text{(S2)}$$

Here, $ZT$ ($T$) function is usually described by using a polymerization function for the numerical calculation. It is noted that error could be generated when $(ZT)_{avg}$ is used to calculate the leg efficiency through the Eq. S1. Alternatively, a more accurate scale, i.e. the engineering figure-of-merit $(\mathrm{ZT})_{\mathrm{eng}}$ and the generic leg efficiency calculation Eq. S3-S4, was introduced by Kim *et al.* by considered temperature dependent Seebeck coefficient $S(T)$, electrical resistivity $\rho(T)$ and thermal conductivity $\kappa(\mathrm{T})$ as following:

$$(ZT)_{eng} = \frac{\left(\int_{T_c}^{T_h} S(T)dT\right)^2}{\int_{T_c}^{T_h} \rho(T)dT \int_{T_c}^{T_h} \kappa(T)dT}(T_h - T_c) \quad \text{(S3)}$$

$$\eta = \eta_c \frac{\sqrt{1+(ZT)_{eng}\, a_1\eta_c^{-1}} - 1}{a_0\sqrt{1+(ZT)_{eng}\, a_1\eta_c^{-1}} + a_2}$$

$$a_i = \frac{S(T_h)\Delta T}{\int_{T_c}^{T_h} S(T)dT} - \frac{\int_{T_c}^{T_h} \tau(T)dT}{\int_{T_c}^{T_h} S(T)dT} W_T\eta_c - iW_J\eta_c \quad \text{(S4)}$$

Here, τ(T) is temperature dependent Thomson coefficient. The definition of $W_T$ and $W_J$ could be found in reference work. When the constant property model is considered, the Eq. S4 could return to the Eq. S1 by applying $W_J$ = 1/2 and τ = 0.

**Supplementary Note II. Specific heat**

The temperature dependent specific heat of $Mg_{3.2}Sb_xBi_{1.99-x}Te_{0.01}$: Mn=0.01 (x=1.2, 1.5) were measured by differential scanning calorimetry (DSC 404C; Netzsch), was used as the reference, which could expressed by a polynomal function, i.e. $C_p=AT^3+BT^2+CT+D$ where T is temperature with a unit of [°C], and the papramers A, B, C and D is given in Table S3.

The specific heat of other $Mg_{3.2}Sb_xBi_{1.99-x}Te_{0.01}$: $Mn_{0.01}$ (x=1.4, 1.3 and 1.1) was estimated by considering a correction of the paprameter D in the polynomal function according to the Dulong–Petit law, and their corressponding paprameters of A, B, C and D were also given in the Table S3. The finally temperature dependent specific heat values were also plotted and shown in the Figure S2.

**Supplementary References**